\documentclass[12pt,epsf]{article}

\usepackage{times}
\usepackage{graphicx}
\usepackage{amsfonts}
\usepackage{amsmath}
\usepackage{amssymb}
\usepackage{amstext}
\usepackage{amsthm}

\usepackage{color}

\usepackage{epsfig}
\psfigdriver{dvips}
\usepackage{latexsym}
\usepackage[matrix,curve]{xypic}
\usepackage{rotating}
\rotdriver{dvips}

\begin{document}

\baselineskip 6mm
\renewcommand{\thefootnote}{\fnsymbol{footnote}}


\newcommand{\nc}{\newcommand}
\newcommand{\rnc}{\renewcommand}


\rnc{\baselinestretch}{1.24}    
\setlength{\jot}{6pt}       
\rnc{\arraystretch}{1.24}   

\makeatletter
\rnc{\theequation}{\thesection.\arabic{equation}}
\@addtoreset{equation}{section}
\makeatother



\nc{\be}{\begin{equation}}

\nc{\ee}{\end{equation}}

\nc{\bea}{\begin{eqnarray}}

\nc{\eea}{\end{eqnarray}}

\nc{\xx}{\nonumber\\}

\nc{\ct}{\cite}

\nc{\la}{\label}

\nc{\eq}[1]{(\ref{#1})}

\nc{\newcaption}[1]{\centerline{\parbox{6in}{\caption{#1}}}}

\nc{\fig}[3]{

\begin{figure}
\centerline{\epsfxsize=#1\epsfbox{#2.eps}}
\newcaption{#3. \label{#2}}
\end{figure}
}


\def\CA{{\cal A}}
\def\CC{{\cal C}}
\def\CD{{\cal D}}
\def\CE{{\cal E}}
\def\CF{{\cal F}}
\def\CG{{\cal G}}
\def\CH{{\cal H}}
\def\CK{{\cal K}}
\def\CL{{\cal L}}
\def\CM{{\cal M}}
\def\CN{{\cal N}}
\def\CO{{\cal O}}
\def\CP{{\cal P}}
\def\CR{{\cal R}}
\def\CS{{\cal S}}
\def\CU{{\cal U}}
\def\CV{{\cal V}}
\def\CW{{\cal W}}
\def\CY{{\cal Y}}
\def\CZ{{\cal Z}}


\def\IB{{\hbox{{\rm I}\kern-.2em\hbox{\rm B}}}}
\def\IC{\,\,{\hbox{{\rm I}\kern-.50em\hbox{\bf C}}}}
\def\ID{{\hbox{{\rm I}\kern-.2em\hbox{\rm D}}}}
\def\IF{{\hbox{{\rm I}\kern-.2em\hbox{\rm F}}}}
\def\IG{{\hbox{{\rm I}\kern-.4em\hbox{\rm G}}}}
\def\IH{{\hbox{{\rm I}\kern-.2em\hbox{\rm H}}}}
\def\IK{{\hbox{{\rm I}\kern-.2em\hbox{\rm K}}}}
\def\IN{{\hbox{{\rm I}\kern-.2em\hbox{\rm N}}}}
\def\IP{{\hbox{{\rm I}\kern-.2em\hbox{\rm P}}}}
\def\IR{{\hbox{{\rm I}\kern-.2em\hbox{\rm R}}}}
\def\IZ{{\hbox{{\rm Z}\kern-.45em\hbox{\rm Z}}}}


\def\a{\alpha}
\def\b{\beta}
\def\d{\delta}
\def\ep{\epsilon}
\def\ga{\gamma}
\def\k{\kappa}
\def\l{\lambda}
\def\s{\sigma}
\def\t{\theta}
\def\w{\omega}
\def\G{\Gamma}


\def\half{\frac{1}{2}}
\def\dint#1#2{\int\limits_{#1}^{#2}}
\def\goto{\rightarrow}
\def\para{\parallel}
\def\brac#1{\langle #1 \rangle}
\def\curl{\nabla\times}
\def\div{\nabla\cdot}
\def\p{\partial}


\def\Tr{{\rm Tr}}
\def\det{{\rm det}}


\def\vare{\varepsilon}
\def\zbar{\bar{z}}
\def\wbar{\bar{w}}
\def\what#1{\widehat{#1}}


\def\ad{\dot{a}}
\def\bd{\dot{b}}
\def\cd{\dot{c}}
\def\dd{\dot{d}}
\def\so{SO(4)}
\def\bfr{{\bf R}}
\def\bfc{{\bf C}}
\def\bfz{{\bf Z}}

\begin{titlepage}


\hfill\parbox{3.7cm} {KIAS-P09039\\
{\tt arXiv:0908.2809}}

\vspace{15mm}

\begin{center}
{\Large \bf  Emergent Geometry from Quantized Spacetime}

\vspace{10mm}

Hyun Seok Yang $^{a,b}$ \footnote{hsyang@ewha.ac.kr} and
M. Sivakumar $^c$ \footnote{mssp@uohyd.ernet.in} \\[10mm]

$^a$ {\sl School of Physics, Korea Institute for Advanced Study,
Seoul 130-012, Korea} \\
$^b$ {\sl Institute for the Early Universe, Ewha Womans University,
Seoul 120-750, Korea} \\
$^c$ {\sl School of Physics, University of Hyderabad, Hyderabad
500046, India}

\end{center}

\thispagestyle{empty}

\vskip1cm


\centerline{\bf ABSTRACT}
\vskip 4mm
\noindent

We examine the picture of emergent geometry arising from a mass-deformed matrix model.
Because of the mass-deformation, a vacuum
geometry turns out to be a constant curvature spacetime such as
$d$-dimensional sphere and (anti-)de Sitter spaces. We show that the
mass-deformed matrix model giving rise to the constant curvature
spacetime can be derived from the $d$-dimensional Snyder algebra.
The emergent geometry beautifully confirms all the rationale inferred
from the algebraic point of view that the $d$-dimensional Snyder
algebra is equivalent to the Lorentz algebra in $(d+1)$-dimensional
{\it flat} spacetime. For example, a vacuum geometry of the mass-deformed matrix model
is completely described by a $G$-invariant
metric of coset manifolds $G/H$ defined by the Snyder algebra. We
also discuss a nonlinear deformation of the Snyder algebra. \\



\vspace{1cm}

\today

\end{titlepage}

\renewcommand{\thefootnote}{\arabic{footnote}}
\setcounter{footnote}{0}

\section{Introduction}

The wave-particle duality in quantum mechanics is a remarkable
consequence of particle dynamics in quantum phase space defined by
$[x^i, p_k] = i \hbar \delta^i_k$. In a classical world with $\hbar =
0$, the wave and the particle are completely independent with
exclusive properties. But, when $\hbar \neq 0$, the particle phase space
becomes noncommutative (NC). As a result, the particle dynamics in the NC phase space
reveals a novel duality such that the wave and the particle are no longer
exclusive entities but complementary aspects of the same physical
reality. That is, they are unified into a single entity with a dual
nature in the quantum world.

A NC spacetime arises from endowing spacetime with
a symplectic structure $B = \half B_{ab} dy^a \wedge dy^b$ and then
quantizing the spacetime with its Poisson structure $\theta^{ab}
\equiv (B^{-1})^{ab}$, treating it as a quantum phase space
described by
\be \la{nc-spacetime}
[y^a, y^b]_\star = i \theta^{ab}.
\ee
Just as the wave-particle duality emerges in the NC phase space
(quantum mechanics) which has never been observed in classical physics,
the NC spacetime \eq{nc-spacetime} may also introduce a new kind of
duality between physical or mathematical entities. So an interesting
question is what kind of duality arises from the quantization of
spacetime triggered by the $\theta$-deformation
\eq{nc-spacetime}. We will see that it is the gauge/gravity duality
as recently demonstrated in \ct{hsy0,hsy1,hsy2,hsyang}.

The gauge/gravity duality in NC spacetime is realized in the context
of emergent gravity where spacetime geometry emerges as a collective
phenomenon of underlying microscopic degrees of freedom defined by
NC gauge fields. Remarkably the emergent gravity reveals a noble
picture about the origin of spacetime, dubbed as emergent spacetime,
which is radically different from any previous physical theory all
of which describe what happens in a given spacetime. The emergent
gravity has been addressed, according to their methodology, from two
facets of quantum field theories: NC field theories
\ct{hsy1,hsy2,hsyang,nc-emergent-1,nc-emergent-2,hsy-nci,nc-emergent-3} and
large $N$ matrix models
\ct{matrix-emergent-1,ikkt,ikkt-nc,ads-cft,matrix-emergent-2,matrix-emergent-3}. But it
turns out \ct{hsy2,hsyang} that the two approaches are intrinsically
related to each other. In particular, the AdS/CFT correspondence
\ct{ads-cft} has been known as a typical example of the emergent gravity
based on a large $N$ matrix model (or gauge theory) which has been
extensively studied for a decade. Furthermore, the emergent gravity
has also been suggested to resolve the cosmological constant problem
and dark energy \ct{pad,hsy-cc}. Nevertheless, there has been little
understanding about why and when the gravity in higher dimensions
can emerge from some kind of lower dimensional quantum field theory
and what the first (dynamical) principle is for the emergent
spacetime.

The issues for the emergent gravity seem to be more accessible from
the approach based on NC geometry. See a recent review, Ref.\ct{szabo}, for
various issues on emergent gravity. In usual commutative spacetime,
a gauge theory such as the electromagnetism is very different from
the gravity described by general relativity since the former is
based on an internal symmetry while the latter is formulated with
the spacetime symmetry. A remarkable property in the NC spacetime
\eq{nc-spacetime} is that the internal symmetry in gauge theory
turns into the spacetime symmetry. This can be seen from the fact
that translations in NC directions are an inner automorphism of NC
$\star$-algebra $\CA_\theta$, i.e., $e^{ik \cdot y} \star
\widehat{f}(y) \star e^{-ik \cdot y} = \widehat{f}(y +  \theta \cdot k)$
for any $\widehat{f}(y) \in \CA_\theta$ or, in its infinitesimal
form,
\be \la{inner-der}
-i [B_{ab}y^b, \widehat{f}(y)]_\star =  \p_a \widehat{f}(y).
\ee

To be specific, let us consider a $U(1)$ bundle supported on a
symplectic manifold $(M, B)$. Because the symplectic structure $B: TM
\to T^* M$ is nondegenerate at any point $y \in M$, we can invert
this map to obtain the map $\theta \equiv B^{-1}: T^* M \to TM$.
This cosymplectic structure $\theta \in \bigwedge^2 TM$ is called
the Poisson structure of $M$ which defines a Poisson bracket
$\{\cdot,\cdot\}_\theta: C^\infty(M) \times C^\infty(M) \to
C^\infty(M)$. The NC spacetime \eq{nc-spacetime} is then obtained by
quantizing the symplectic manifold $(M, B)$ with the Poisson
structure $\theta = B^{-1}$. An important point is that the gauge
symmetry acting on $U(1)$ gauge fields as $A \to A + d \phi$ is a
diffeomorphism symmetry generated by a vector field $X$ satisfying
$\CL_X B = 0$, which is known as the symplectomorphism in symplectic
geometry. In other words, $U(1)$ gauge transformations are generated
by the Hamiltonian vector field $X_\phi$ satisfying $\iota_{X_\phi}B
+ d \phi =0$ and the action of $X_\phi$ on a smooth function $f(y)
\in C^\infty(M)$ is given by
\be \la{semi-gauge-tr} \delta f(y)
\equiv X_\phi (f)(y) = \{ f, \phi \}_{\theta}(y).
\ee
Therefore the gauge symmetry \eq{semi-gauge-tr} on the symplectic manifold $(M,
B)$ should be regarded as a spacetime symmetry rather than an
internal symmetry \ct{hsy1}.

The above reasoning implies that $U(1)$ gauge fields in NC spacetime
can be realized as a spacetime geometry like as the gravity in
general relativity \ct{hsy1,hsy2,hsyang}. In general relativity the
equivalence principle beautifully explains why the gravitational
force has to manifest itself as a spacetime geometry. If the
gauge/gravity duality is realized in NC spacetime, a natural
question is what is the corresponding equivalence principle for the
geometrization of the electromagnetic force. Because the geometrical
framework of NC spacetime is apparently based on the symplectic
geometry in sharp contrast to the Riemannian geometry, the question
should be addressed in the context of the symplectic geometry rather
than the Riemannian geometry. Remarkably it turns out that NC
spacetime admits a novel form of the equivalence principle such that
there ``always" exists a coordinate transformation to locally
eliminate the electromagnetic force \ct{hsyang}. This geometrization
of the electromagnetism is inherent as an intrinsic property in the
symplectic geometry known as the Darboux theorem or the Moser lemma
\ct{moser}. As a consequence, the electromagnetism in NC spacetime
can be realized as a geometrical property of spacetime like gravity.

This noble form of the equivalence principle can be understood as
follows \ct{hsyang}. The presence of fluctuating gauge fields on a
symplectic manifold $(M,B)$ appears as a deformation of the
symplectic manifold $(M,B)$ such that the resulting symplectic
structure is given by $\omega_1 \equiv B+F$ where $F = dA$. Because
the original symplectic structure $\omega_0 = B$ is a nondegenerate
and closed two-form, the associated map $B^\flat : TM \to T^* M$ is a
vector bundle isomorphism. Therefore there exists a natural pairing
$\Gamma(TM) \to \Gamma(T^* M): X \mapsto B^\flat (X) = \iota_X B$
between $C^\infty$-sections of tangent and cotangent bundles. Because
the $U(1)$ gauge field $A$ on $M$ only appears as the combination
$\omega_1 = B + dA$, one may identify the connection $A$ with an
element in $\Gamma(T^* M)$ such that
\be \la{moser-eq}
\iota_X B + A = 0.
\ee
The identification \eq{moser-eq} is defined up to
symplectomorphisms or equivalently $U(1)$ gauge transformations,
that is, $X \sim X + X_\phi \Leftrightarrow  A \sim A + d \phi$
where $\iota_{X_\phi} B = d \phi$. Using the Cartan's magic formula
$\mathcal{L}_X = d \iota_X + \iota_X d$ and so $[\mathcal{L}_X,
d]=0$, it is easy to see that $\omega_1 = B + dA = B - \mathcal{L}_X
B$ and $d \omega_1 = 0$ because of $dB=0$. This means that a smooth
family $\omega_t = \omega_0 + t(\omega_1 - \omega_0)$ of symplectic
structures joining $\omega_0$ to $\omega_1$ is all
deformation-equivalent and there exists a map $\phi: M \times
\mathbf{R} \to M$ as a flow - a one-parameter family of
diffeomorphisms - generated by the vector field $X_t$ satisfying
$\iota_{X_t} \omega_t + A = 0$ such that $\phi_t^*(\omega_t) =
\omega_0$ for all $0 \leq t \leq 1$.

This can be explicitly checked by considering a local Darboux chart
$(U; y^1, \cdots, y^{2n})$ centered at $p \in M$ and valid on the
neighborhood $U$ such that $\omega_0|_U = \frac{1}{2} B_{ab} dy^a
\wedge dy^b$ where $B_{ab}$ is a constant symplectic matrix of rank
$2n$. Now consider a flow $\phi_t: U \times [0,1] \to M$ generated
by the vector field $X_t$ satisfying \eq{moser-eq}. Under the action
of $\phi_\epsilon$ with an infinitesimal $\epsilon$, one finds that
a point $p \in U$ whose coordinate is $y^a$ is mapped to
$\phi_\epsilon (y) \equiv x^a(y) = y^a + \epsilon X^a(y)$. Using the
inverse map $\phi^{-1}_\epsilon: x^a \mapsto y^a (x)= x^a - \epsilon
X^a(x)$, the symplectic structure $\omega_0|_U = \frac{1}{2} B_{ab}
(y) dy^a \wedge dy^b$ can be expressed as
\bea \la{moser}
(\phi^{-1}_\epsilon)^* (\omega_0|_y) &=& \frac{1}{2} B_{ab}(x -
\epsilon X) d(x^a - \epsilon X^a) \wedge d(x^b - \epsilon X^b) \xx
& \approx & \frac{1}{2} \Big[ B_{ab} - \epsilon X^\mu (\partial_\mu
B_{ab} + \partial_b B_{\mu a} + \partial_a B_{b \mu}) + \epsilon
\Big( \partial_a (B_{b\mu} X^\mu) - \partial_b (B_{a \mu} X^\mu)
\Big) \Big] dx^a \wedge dx^b \xx &\equiv& B + \epsilon F
\eea
where $A_a(x) = B_{a\mu}(x) X^\mu(x)$ or $\iota_X B + A = 0$ and $dB = 0$
was used for the vanishing of the second term. Equation \eq{moser} can be
rewritten as $\phi_\epsilon^* (B + \epsilon F) = B$, which means
that the electromagnetic force $F = dA$ can always be eliminated by
a local coordinate transformation generated by the vector field $X$
satisfying Eq.\eq{moser-eq}.

Surprisingly it is easy to understand how the Darboux theorem in
symplectic geometry manifests itself as a novel form of the
equivalence principle such that the electromagnetism in NC spacetime
can be regarded as a theory of gravity \ct{hsy1,hsy2,hsyang}. It is
well known that, for a given Poisson algebra $(C^\infty(M),
\{\cdot,\cdot\}_\theta)$, there exists a natural map $C^\infty (M)
\to TM: f \mapsto X_f$ between smooth functions in $C^\infty(M)$ and
vector fields in $TM$ such that
\be \la{duality}
X_f (g) = \{g, f\}_{\theta}
\ee
for any $g \in C^\infty(M)$. Indeed the assignment
\eq{duality} between a Hamiltonian function $f$ and the
corresponding Hamiltonian vector field $X_f$ is the Lie algebra
homomorphism in the sense
\be \la{poisson-lie}
X_{\{f,g\}_{\theta}} = - [X_f,X_g]
\ee
where the right-hand side represents the Lie bracket
between the Hamiltonian vector fields.

The correspondence \eq{duality} between the Poisson algebra
$(C^\infty(M), \{\cdot,\cdot\}_\theta)$ and vector fields in
$\Gamma(TM)$ can be generalized to the NC $\star$-algebra
$(\CA_\theta, [\cdot,\cdot]_\star)$ by considering an adjoint
operation of NC gauge fields $\widehat{D}_a(y) \in \CA_\theta$ as follows
\bea \la{nc-vector}
ad_{\widehat{D}_a} [\widehat{f}] (y)
&\equiv & - i [ \widehat{D}_a(y), \widehat{f}(y) ]_\star = -
\theta^{\mu \nu} \frac{\partial D_a(y)}{\partial y^\nu}
\frac{\partial f(y)}{\partial y^\mu} + \cdots \xx
&\equiv & V_a[f](y) + {\cal O}(\theta^3).
\eea
The leading term in Eq.\eq{nc-vector} exactly recovers the vector fields in
Eq.\eq{duality} and the vector field $V_a (y) = V_a^\mu (y)
\frac{\partial}{\partial y^\mu} \in \Gamma(TM_y)$ takes values in
the Lie algebra of volume-preserving diffeomorphisms since $\p_\mu
V_a^\mu = 0$ by definition. But it can be shown \ct{hsyang} that the
vector fields $V_a \in \Gamma(TM)$ are related to the orthonormal
frames (vielbeins) $E_a$ by $V_a = \lambda E_a$ where $\lambda^2 =
\det V_a^\mu$. Therefore, we see that the Darboux theorem in
symplectic geometry implements a deep principle to realize a
Riemannian manifold as an emergent geometry from NC gauge fields
through the correspondence \eq{nc-vector} whose metric is given by \ct{hsy2,hsyang}
\be \la{emergent-metric}
ds^2 = g_{ab} E^a \otimes E^b = \lambda^2 g_{ab} V^a_\mu V^b_\nu dy^\mu \otimes dy^\nu
\ee
where $E^a = \lambda V^a \in \Gamma(T^*M)$ are dual oneforms.

If a coordinate transformation is generated by a Hamiltonian vector
field $X_\phi$ satisfying $\iota_{X_\phi} B = d \phi$ or $X^\mu_\phi
= \theta^{\mu\nu} \partial_\nu \phi$, the symplectic structure
remains intact as can easily be checked from Eq.\eq{moser}. It
should be the case since the symplectomorphism generated by the
Hamiltonian vector field is equal to the $U(1)$ gauge
transformation. So let us look at a response of the metric
\eq{emergent-metric} under the coordinate transformation in the
symplectomorphism or the $U(1)$ gauge transformation. Using the
definition of the vector fields in Eq.\eq{nc-vector}, one can
rewrite the inverse metric of Eq.\eq{emergent-metric} as follows
\bea \la{inverse-emergent-metric}
\Big(\frac{\partial}{\partial s}
\Big)^2 &=& g^{ab} E_a \otimes E_b = \lambda^{-2} g^{ab} V_a^\mu
V_b^\nu \partial_\mu \otimes
\partial_\nu \equiv \mathfrak{G}^{\mu\nu} \partial_\mu \otimes
\partial_\nu \xx
&=& \theta^{\mu a} \theta^{\nu b} G_{\alpha\beta} \frac{\partial
x^\alpha}{\partial y^a} \frac{\partial x^\beta}{\partial y^b}
\partial_\mu \otimes \partial_\nu \xx
&=& \theta^{\mu a} \theta^{\nu b} (G_{ab} + \CL_{X_\phi} G_{ab})
\partial_\mu \otimes \partial_\nu
\eea
where $G_{ab} = - \lambda^{-2} B_{ac} g^{cd} B_{db}$ and
$x^\alpha(y) = y^\alpha + X^\alpha_\phi (y)$. For consistency the metric
\eq{inverse-emergent-metric} should remain intact under the $U(1)$
gauge transformation or the symplectomorphism since it does not
change the symplectic structure. It is easy to see that this
consistency condition is equivalent to require $\CL_{X_\phi}
G_{ab}= 0$ since $V_a^\mu = \delta^\mu_a$ in this case and so
$\lambda^2 = \det V_a^\mu = 1$. Therefore, we get a consistent result
that the $U(1)$ gauge transformation or the symplectomorphism
corresponds to a Killing symmetry and the emergent metric
\eq{emergent-metric} does not change, i.e., $\mathfrak{G}_{\mu\nu} =
g_{\mu\nu}$.

As emphasized by Elvang and Polchinski \ct{polchinski}, the
emergence of gravity requires the emergence of spacetime itself.
That is, spacetime is not given {\it a priori} but defined by
``spacetime atoms", NC gauge fields in our case, in quantum gravity
theory. It should be required for consistency that the entire
spacetime including a flat spacetime has to be emergent from NC
gauge fields. In other words, the emergent gravity should
necessarily be background independent where any spacetime structure
is not {\it a priori} assumed but defined from the theory. Let us
elucidate using the relation between a matrix model and a NC gauge
theory \ct{ikkt-nc,sw,seiberg} how the emergent gravity based on the
NC geometry achieves the background independence \ct{hsy2,hsyang}.

Consider the zero-dimensional IKKT matrix model \ct{ikkt} whose action
is given by
\be \la{ikkt}
S_{IKKT} =  - \frac{1}{4} \Tr \big( [X_a, X_b][X^a, X^b] \big).
\ee
Because the action \eq{ikkt} is zero-dimensional, it does not assume the prior existence of any
spacetime structure. There are only a bunch of $N \times N$
Hermitian matrices $X^a \; (a=1, \cdots, 2n)$ which are subject to a
couple of algebraic relations given by
\bea \la{matrix-eom}
&& [X_a, [X^a, X^b]] = 0, \\
\la{matrix-bianchi}
&& [X^a, [X^b, X^c]] + [X^b, [X^c, X^a]] + [X^c, [X^a, X^b]] = 0.
\eea

In order to consider fluctuations around a vacuum of the matrix
theory \eq{ikkt}, first one has to specify the vacuum of the theory
where all fluctuations are supported. Of course, the vacuum solution
itself should also satisfy the Eqs. \eq{matrix-eom} and
\eq{matrix-bianchi}. Suppose that the vacuum solution is given by
$X^a_{{\rm vac}} = y^a$. In the limit $N \to \infty$, the Moyal NC
space defined by Eq.\eq{nc-spacetime} where $\theta^{ab}$ is a
constant matrix of rank $2n$ definitely satisfies the equations of
motion \eq{matrix-eom} as well as the Jacobi identity
\eq{matrix-bianchi}. Furthermore, in this case, the matrix algebra
$(M_N, [\cdot,\cdot])$ defining the action \eq{ikkt} can be mapped
to the NC $\star$-algebra $(\CA_\theta, [\cdot,\cdot]_\star)$
defined by the NC space \eq{nc-spacetime} \ct{ikkt-nc}. To be
explicit, let us expand the large $N$ matrices $X^a \equiv
\theta^{ab} \widehat{D}_b$ around the Moyal vacuum \eq{nc-spacetime}
as follows:
\be \la{x-exp}
\widehat{D}_a(y) = B_{ab} y^b + \widehat{A}_a(y).
\ee
Note that
\bea \la{nc-curvature}
-i [\widehat{D}_a (y), \widehat{D}_b (y)]_\star
&=& \partial_a \widehat{A}_b(y) - \partial_b \widehat{A}_a (y)
- i [\widehat{A}_a (y), \widehat{A}_b (y)]_\star - B_{ab} \xx
&=& \widehat{F}_{ab}(y) - B_{ab}.
\eea
Then the IKKT matrix model \eq{ikkt} becomes the NC $U(1)$ gauge theory in
$2n$ dimensions \ct{ikkt-nc,seiberg}
\be \la{ncu1}
\widehat{S}_{NC} =  \frac{1}{4g_{YM}^2} \int d^{2n} y  G^{ac} G^{bd} \big(
\widehat{F} - B \big)_{ab} \star \big( \widehat{F} - B \big)_{cd}
\ee
where $G^{ab} = \theta^{ac} \theta^{bc}$ and $\Tr \to \int
\frac{d^{2n} y}{(2\pi)^n |{\rm Pf} \theta|}$ and we have recovered a
$2n$-dimensional gauge coupling constant $g_{YM}^2$ \ct{sw}.

According to the correspondence \eq{nc-vector}, the NC gauge fields
$\widehat{D}_a(y) \in \CA_\theta$ in Eq.\eq{x-exp} are mapped to
(generalized) vector fields $\widehat{V}_a(y) \equiv
ad_{\widehat{D}_a} (y)$ as an inner derivation in $\CA_\theta$
\ct{hsy1,hsy2,hsyang}. In particular, we have the property
\be \la{jacobi-derivation}
[ ad_{\widehat{D}_a}, ad_{\widehat{D}_b} ]_\star = ad_{\widehat{F}_{ab}}
= [\widehat{V}_a, \widehat{V}_b]_\star
\ee
where $[\widehat{V}_a, \widehat{V}_b]_\star = [V_a, V_b] + \CO(\theta^3)$ is a generalization
of the Lie bracket to the generalized vector fields in Eq.\eq{nc-vector}. Using the maps
in Eqs.\eq{nc-vector} and \eq{jacobi-derivation}, one can further
deduce that
\be \la{jacobi-3}
[ ad_{\widehat{D}_a}, [ ad_{\widehat{D}_b}, ad_{\widehat{D}_c} ]_\star ]_\star =
ad_{\widehat{D}_a \widehat{F}_{bc}} = [\widehat{V}_a, [\widehat{V}_b, \widehat{V}_c]_\star ]_\star.
\ee
Using the relation \eq{jacobi-3}, one can easily show that the equations of motion for
NC gauge fields derived from the action \eq{ncu1} are mapped to the
geometric equations for (generalized) vector fields defined by
Eq.\eq{nc-vector} \ct{hsyang}:
\bea \la{nc-jacobi}
&& \widehat{D}_{[a} \widehat{F}_{bc]} = 0 \quad \Leftrightarrow \quad
 [\widehat{V}_{[a}, [\widehat{V}_b,
\widehat{V}_{c]}]_\star ]_\star = 0, \\
\la{nc-eom}
&&  \widehat{D}^{a} \widehat{F}_{ab} = 0 \quad \Leftrightarrow \quad
[\widehat{V}^a, [\widehat{V}_a, \widehat{V}_b]_\star ]_\star = 0.
\eea
To be specific, if one confines to the leading order in Eq.\eq{nc-vector} where one
recovers usual vector fields, the Jacobi identity
\eq{matrix-bianchi} [or the Bianchi identity \eq{nc-jacobi} for NC
gauge fields] is equivalent to the first Bianchi identity for
Riemann tensors, i.e., $R_{[abc]d} = 0$ and the equations of motion
\eq{matrix-eom} for $N \times N$ matrices or \eq{nc-eom} for NC
gauge fields are mapped to the Einstein equations, $R_{ab} -
\frac{1}{2} g_{ab} R = 8 \pi G T_{ab}$, for the emergent metric
\eq{emergent-metric} \ct{hsyang}.

Though the emergence of Einstein gravity from NC gauge fields is
shown after some non-trivial technical computations \ct{hsyang}, it
can easily be verified for the self-dual sector without any further
computation. First notice the following equality directly derived
from Eq.\eq{jacobi-derivation}
\be \la{nc-gr-instanton}
\widehat{F}_{ab} = \pm \frac{1}{2} {\varepsilon_{ab}}^{cd}
\widehat{F}_{cd} \quad \cong  \quad [\widehat{V}_a,
\widehat{V}_b]_\star = \pm \frac{1}{2} {\varepsilon_{ab}}^{cd}
[\widehat{V}_c, \widehat{V}_d]_\star.
\ee
Because $[\widehat{V}_a, \widehat{V}_b]_\star = [V_a, V_b] + \CO(\theta^3)$, the right-hand
side of Eq.\eq{nc-gr-instanton} in commutative, i.e. $\CO(\theta)$,
limit describes self-dual and Ricci-flat four-manifolds as was
rigorously proved in \ct{hsy0,hsyang,gr-nc-instanton}. In other
words, the self-dual Einstein gravity arises from the leading order
of self-dual NC gauge fields \ct{hsy-nci}.

One can trace the emergent metric \eq{emergent-metric} back to see
where the flat spacetime comes from. It turns out \ct{hsy-cc} that
the flat spacetime is emergent from the uniform condensation of
gauge fields giving rise to the NC spacetime \eq{nc-spacetime}. This
is a tangible difference from Einstein gravity where the flat
spacetime is a completely empty space. Furthermore, since gravity
emerges from NC gauge fields, the parameters, $g^2_{YM}$ and
$|\theta|$, defining a NC gauge theory should be related to the
Newton constant $G$ in emergent gravity. A simple dimensional
analysis shows that $\frac{G \hbar^2}{c^2} \sim g^2_{YM}|\theta|$.
In four dimensions, this relation immediately leads to the fact that
the energy density of the vacuum \eq{nc-spacetime} is $\rho_{{\rm
vac}} \sim |B_{ab}|^2 \sim M^4_P$ where $M_P = (8\pi G)^{-1/2} \sim
10^{18} GeV$ is the Planck mass. Therefore the emergent gravity
reveals a remarkable picture that the huge Planck energy $M_P$ is
actually used to generate a flat spacetime. It is very surprising but
should be expected from the background independence of the emergent
gravity that a flat spacetime is not free gratis but a result of
Planck energy condensation in vacuum. Hence the vacuum energy does
not gravitate unlike Einstein gravity. It was argued in
\ct{hsyang,hsy-cc} that this emergent spacetime picture will be
essential to resolving the cosmological constant problem, to
understanding the nature of dark energy and to explaining why gravity is
so weak compared to other forces.

In this paper we will generalize the picture of emergent geometry to
the case with a nontrivial vacuum geometry, especially, a constant
curvature spacetime. This kind of emergent geometry will arise from
a mass-deformed matrix model. The subsequent parts of this paper
will be organized as follows.

In Sec. 2, we will consider the matrix model of $SO(3-p, p)$ Lie
algebra with $p=0,1,2$ which is the matrix version of
Maxwell-Chern-Simons theory or massive Chern-Simons theory
\ct{deser-jackiw}. We show that either compact or non-compact
(fuzzy) Riemann surfaces such as a two-dimensional sphere and
(anti-)de Sitter spaces are emergent from the matrix model. A
well-known example of quantized compact Riemann surfaces is a fuzzy
sphere \ct{fuzzys2}. We discuss how a nonlinear deformation of the
underlying Lie algebra can trigger a topology change of the Riemann
surfaces \ct{fuzzy-riemann}.

In Sec. 3, we will generalize the matrix model of two-dimensional
Riemann surfaces to higher dimensions. The emergent geometry in
higher dimensions is deduced from a mass-deformed IKKT matrix model
\ct{mass-matrix}. Because of the mass deformation, a vacuum geometry is
no longer flat but a constant curvature spacetime such as
a $d$-dimensional sphere and (anti-)de Sitter spaces. We show that the
mass-deformed matrix model giving rise to the constant curvature
spacetime can be derived from the $d$-dimensional Snyder algebra
\ct{snyder}. The emergent gravity beautifully confirms all the
rationale inferred from the algebraic point of view that the
$d$-dimensional Snyder algebra is equivalent to the Lorentz algebra
in $(d+1)$-dimensional {\it flat} spacetime. We also discuss a
nonlinear deformation of the Snyder algebra.

In Sec. 4, we show that a vacuum geometry of the mass-deformed
matrix model is completely described by a $G$-invariant metric of
coset manifolds $G/H$ \ct{coset} defined by the Snyder algebra. We
thus advocate the picture that the geometrical aspects of emergent
gravity for the mass-deformed matrix model can be nicely captured by
the equivalence between the $d$-dimensional Snyder algebra and the
$(d+1)$-dimensional Lorentz algebra. Finally we conclude with
several remarks about the significance of emergent geometry based on
the results we have obtained.

In the Appendix, it is shown that the two-dimensional Snyder algebra is
precisely equal to the three-dimensional $SO(3-p,p)$
Lie algebra in Sec. 2.

\section{Two-dimensional Manifolds from Matrix Model}

Consider the following master matrix action:
\begin{equation} \la{master-action}
S_M = \Tr \Big( \frac{g^2_{YM}}{2} P_A P^A -
\lambda P_A X^A + \frac{i \kappa}{3!} \varepsilon_{ABC} X^A [X^B, X^C] \Big)
\ee
where $\lambda = \kappa g_{YM}^2$ and $A,B, \cdots = 1,2,3$. The
equations of motion are read as
\bea \la{eom-p}
&& P_A =  \frac{i}{2g^2_{YM}} \varepsilon_{ABC} [X^B, X^C], \\
\la{eom-x}
&& P^A = \kappa X^A.
\eea

Substituting Eq.\eq{eom-p} into the master action \eq{master-action}
leads to the matrix version of Maxwell-Chern-Simons action
\ct{matrix-emergent-3}
\be \la{maxcs}
S_{MCS} = - \frac{1}{g^2_{YM}} \Tr \Big(\frac{1}{4}[X^A, X^B]^2 +
\frac{i \lambda}{3} \varepsilon_{ABC} X^A [X^B, X^C] \Big)
\ee
while Eq.\eq{eom-x} leads to the matrix version of massive
Chern-Simons theory
\be \la{mass-cs}
S_{mCS} = \kappa \Tr \Big(\frac{i}{3!} \varepsilon_{ABC} X^A [X^B,
X^C] - \frac{\lambda}{2} X_A X^A  \Big).
\ee
Thus we establish the matrix version of the duality between
topologically massive electrodynamics and self-dual massive model
\ct{deser-jackiw}. Therefore, it is enough to solve either
Eq.\eq{maxcs} or Eq.\eq{mass-cs} to get physical spectra.

From the action \eq{mass-cs}, one can see that the equations of
motion are given by the $SO(3-p, p)$ Lie algebra with $p=0,1,2$
\be
\la{eom-su2} [X^A , X^B] = - i \lambda {\varepsilon^{AB}}_C X^C.
\ee
We are interested in deriving a two-dimensional manifold from the
Lie algebra \eq{eom-su2} where the Casimir invariant is given by
\footnote{\label{s2s3}It is well known that the Lie algebra
\eq{eom-su2} can be represented by differential operators as tangent
vectors on some manifold, which is actually the result we want to
realize using the map \eq{nc-vector}. Without imposing the Casimir
invariant \eq{casimir}, one gets a three-dimensional manifold, e.g.,
${\bf S}^3$ from $SU(2)$ algebra. In our case, imposing
Eq.\eq{casimir}, we will get a two-dimensional manifold instead. As
will be discussed in the Appendix, the $SO(3-p, p)$ Lie algebra in
Eq.\eq{eom-su2} will then be interpreted as the Lorentz algebra of
an ambient three-dimensional space, which is precisely the three-dimensional
version of Eq.\eq{high-lorentz}.}
\be \la{casimir}
 g_{AB} X^A X^B \equiv (-)^\sharp R^2.
\ee
We will consider three cases depending on the choice of metric
$g_{AB}$: (I) $g_{AB} = {\rm diag}(1,1,1)$ with $\sharp =0$, (II)
$g_{AB} = {\rm diag}(-1,1,1)$ with $\sharp = 0$, and (III) $g_{AB} =
{\rm diag}(-1,1,-1)$ with $\sharp = 1$. They describe a
two-dimensional manifold $M$ of radius $R$ given by Eq.\eq{casimir}
in the classical limit: (I) sphere ${\bf S}^2$, (II) de Sitter space
$dS_2$, and (III) anti-de Sitter space $AdS_2$, which may be
represented by the cosets $SO(3)/SO(2)$, $SO(2,1)/SO(1,1)$, and
$SO(1,2)/SO(1,1)$, respectively. See Sec. 4 for the coset space
realization of two-dimensional hypersurface $M$.

We will first clarify how the Lie algebra \eq{eom-su2} arises from
the quantization of two-dimensional (orientable) manifolds
\ct{fuzzy-riemann,fuzzy-ads}. Let $M$ be an orientable two-manifold and
$\omega \in \Omega^2(M)$ a volume form. Then $\omega$ is
nondegenerate (since $\omega \neq 0$ everywhere) and obviously
closed, i.e., $d\omega = 0$. Therefore, any orientable two-manifold $M$
is a symplectic manifold. A unique feature in two dimensions is that
a symplectic two-form is just a volume form. Hence any two volume
forms $\omega$ and $\omega'$ on a two-dimensional manifold $M$,
defining the same orientation and having the same total volume, will
be related by an exact two-form; $\omega' =
\omega + dA$. This is a well-known result on volume forms due to
Moser \ct{moser}. (For a noncompact manifold, we would need to
introduce a compact support of symplectic form.) In particular,
every closed symplectic two-manifold is determined up to local
isotopic deformations by its genus and total volume. This implies
that a nontrivial deformation of two-dimensional manifolds will be
encoded only in volume and topology changes up to volume-preserving
metric (shape) deformations. We will see that this feature still
persists in a two-dimensional NC manifold.

To begin with, let us introduce a local Darboux chart $(U; y^1,
y^2)$ centered at $p \in M$ and valid on a neighborhood $U$ such
that $\omega|_U = \half B_{ab} dy^a \wedge dy^b = - dy^1 \wedge
dy^2$. The Poisson bracket for $f,g \in C^\infty(M)$ is then defined
in terms of local coordinates $y^a \; (a=1,2)$
\be \la{2-poisson}
\{f, g\}_\theta = \theta^{ab} \frac{\partial f}{\partial y^a}
\frac{\partial g}{\partial y^b}
\ee
where $\theta^{12} = 1$. We will consider the two-dimensional
manifold $M$ as a hypersurface embedded in $\IR^{3-p, p}$ and
described by $L^A = L^A(y), \; A=1,2,3$, satisfying the relation
\eq{casimir}. For example, one can choose $y^a = (\cos
\theta, \varphi)$ for ${\bf S}^2$, $y^a = (\sinh t,
\varphi)$ for $dS_2$, and $y^a = (t, \sinh x)$ for $AdS_2$ as
follows.

(I) ${\bf S^2}$ of unit radius:
\be \la{emb-sphere}
L^1  = \sqrt{1 - y^2} \cos \varphi, \quad L^2  = \sqrt{1 - y^2}
\sin  \varphi,
\quad L^3 = y,
\ee
where $y = \cos \theta$.

(II) $dS_2$ of unit radius:
\be \la{emb-ds2}
L^1 = - y, \quad L^2  = \sqrt{1 + y^2} \sin \varphi, \quad L^3 =
\sqrt{1 + y^2} \cos \varphi,
\ee
where $y = \sinh t$.

(III) $AdS_2$ of unit radius:
\be \la{emb-ads2}
L^1 = \sqrt{1 + y^2} \cos t, \quad L^2 = - y, \quad L^3  =
\sqrt{1 + y^2} \sin t,
\ee
where $y = \sinh x$.

It is easy to see that the above coordinate system $L^A(y) \in
C^\infty(M)$ satisfies a linear Poisson structure under the Poisson
bracket \eq{2-poisson}
\be \la{lie-poisson}
\{L^A , L^B\}_\theta = - {\varepsilon^{AB}}_C L^C.
\ee
The coordinate system $L_{A}(y) = g_{AB} L^B(y) \in C^\infty(M)$
satisfying the constraint \eq{casimir} can be mapped to vector
fields $V_A^{(0)} (y) = V_A^{(0)a}(y)
\frac{\partial }{\partial y^a}
\in \Gamma(TM)$ according to Eq.\eq{duality} as
\be \la{drei-bein}
V_A^{(0)} = \theta^{ab} \frac{\partial L_{A}}{\partial
y^b}\frac{\partial }{\partial y^a}.
\ee
The two-dimensional metric on $M$ is then determined by the vector
fields \eq{drei-bein} where the inverse metric is given by
\be \la{2-metric}
\mathfrak{G}_{(0)}^{ab} = (\det \mathfrak{G}^{(0)}_{ab})^{-1}
g^{AB} V_A^{(0)a} V_B^{(0)b}
\ee
and so the two-dimensional (emergent) metric reads as
\be \la{2-emetric}
\mathfrak{G}^{(0)}_{ab} =
g_{AB} \frac{\partial L^{A}}{\partial y^a} \frac{\partial
L^{B}}{\partial y^b}.
\ee
One can easily check that the resulting metric $ds^2 =
\mathfrak{G}^{(0)}_{ab} dy^a dy^b$ is equivalent to the induced
metric from the standard flat metric $ds^2 = g_{AB} dL^A dL^B$ on
$\IR^{3-p, p}$:

\bea \la{two-metric}
{\rm (I)}: \;\; ds^2 &=& \frac{dy^2}{1-y^2} + (1-y^2) d\varphi^2
\xx &=& d\theta^2 + \sin^2 \theta d \varphi^2, \\
{\rm (II)}: \;\; ds^2 &=& - \frac{dy^2}{1+y^2} + (1+ y^2) d\varphi^2
\xx &=& - dt^2 + \cosh^2 t d \varphi^2, \\
{\rm (III)}: \;\; ds^2 &=& -  (1+ y^2) dt^2 + \frac{dy^2}{1+y^2} \xx
&=& - \cosh^2 x dt^2 + dx^2.
\eea

As it should be, we see here that the metric \eq{2-emetric}
determined by the vector fields in Eq.\eq{drei-bein} is just the
induced metric on a two-dimensional surface $M$ embedded in
$\IR^{3-p, p}$. Let us now consider a generic fluctuation of the
surface $M$ around the vacuum geometry (I)-(III) described by
\be \la{fluctuation}
X^A(y) = L^A(y) + A^A(y).
\ee
The fluctuating coordinate system \eq{fluctuation} satisfies the
following Poisson bracket relation
\be \la{poisson-xx}
\{X^A, X^B\}_\theta  = - {\varepsilon^{AB}}_C X^C + F^{AB}
\ee
where
\be \la{poisson-field}
F^{AB} = \{L^A, A^B\}_\theta - \{L^B, A^A\}_\theta +
\{ A^A, A^B \}_\theta + {\varepsilon^{AB}}_C A^C.
\ee
Note that the field strength $F^{AB}$ in Eq.\eq{poisson-xx} cannot
be arbitrary since the Poisson algebra \eq{poisson-xx} should
satisfy the Jacobi identity, $\varepsilon_{ABC} \{X^A, \{X^B,
X^C\}_\theta \}_\theta = 0$. This constraint can be solved by taking
the field strength $F^{AB}$ in Eq.\eq{poisson-xx} as the form
\be \la{field-form}
F^{AB}(X)= \varepsilon^{ABC} \frac{\partial F(X)}{\partial X^C}
\ee
with an arbitrary smooth function $F(X)$ defined in ${\cal M} =
\IR^{3-p,p}$ because we have
$$ \half \varepsilon_{ABC} \{X^A, \{X^B,
X^C\}_\theta \}_\theta =  \{X^A,  \frac{\partial F(X)}{\partial X^A}
\}_\theta = \frac{\partial^2 F(X)}{\partial X^A \partial X^B} \{X^A,
X^B \}_\theta = 0. $$ Then the Poisson bracket relation
\eq{poisson-xx} can be written as follows
\be \la{gen-poisson}
\{X^A, X^B\}_\theta  = \varepsilon^{ABC} \frac{\partial G(X)}{\partial X^C}
\ee
where the polynomial $G(X)$ is defined in ${\cal M} =
\IR^{3-p,p}$ and given by
\be \la{poly-g}
G(X) = F(X) - \half g_{AB} X^A X^B + \rho.
\ee

It is interesting to notice that, for $f, g \in C^\infty(M)$,
\bea \la{nambu-poisson}
\{f(X), g(X)\}_\theta & = & \frac{\partial f(X)}{\partial X^A}
\frac{\partial g(X)}{\partial X^B} \{X^A, X^B\}_\theta =
\varepsilon^{ABC} \frac{\partial G(X)}{\partial X^A} \frac{\partial f(X)}{\partial X^B}
\frac{\partial g(X)}{\partial X^C} \xx
&\equiv & \{G(X), f(X), g(X)\}_{NP}
\eea
where $\{f(X), g(X), h(X) \}_{NP}$ is the Nambu-Poisson bracket for
arbitrary functions $f, g, h \in C^\infty({\cal M})$. The
Nambu-Poisson bracket satisfies some fundamental identity (see
Eq.(3.2) in
\ct{pmho})
\bea \la{fund-id}
\{f_1, f_2, \{f_3, f_4, f_5 \}_{NP} \}_{NP}
&=& \{\{f_1, f_2, f_3 \}_{NP}, f_4, f_5 \}_{NP} +  \{ f_3, \{f_1,
f_2, f_4 \}_{NP}, f_5 \}_{NP} \xx && + \{f_3, f_4, \{f_1, f_2, f_5
\}_{NP} \}_{NP}.
\eea
Then one can easily see that the Jacobi identity for the Poisson
bracket \eq{gen-poisson} is actually the statement of the
fundamental identity
\eq{fund-id} since
\bea \la{jacobi-fund-id}
&& \{f, \{g, h\}_\theta \}_\theta + \{g, \{h, f\}_\theta \}_\theta +
\{h, \{f, g\}_\theta \}_\theta \xx
&=& \{G, f, \{G, g, h \}_{NP} \}_{NP} + \{G, g, \{G, h, f \}_{NP}
\}_{NP} + \{G, h, \{G, f, g \}_{NP} \}_{NP} \xx
&=& - \{\{f, G, G \}_{NP}, g, h \}_{NP} = 0.
\eea

In order to allow a general fluctuation including topology and
volume changes of the two-dimensional surface $M$, suppose that the
function $F(X)$ in Eq.\eq{field-form} is an arbitrary polynomial in
three variables in ${\cal M} = \IR^{3-p,p}$. The two-dimensional surface
$M$ embedded in ${\cal M}$ will be defined by zeros of the
polynomial
\eq{poly-g}, i.e., $M = G^{-1}(\{0\})$ and $X^A(y)$ in
Eq.\eq{fluctuation} will be a local parameterization of $M$ in terms
of Darboux coordinates $y^a$. For example, a Riemann surface
$\Sigma_g$ of genus $g$ is described by
\be \la{riemann-poly}
G(\vec{x}) = (P(x) + y^2)^2 + z^2 - \mu^2, \quad \vec{x} =(x,y,z)
\in \IR^3,
\ee
with the polynomial $P(x) = x^{2k} + a_{2k-1} x^{2k-1} + \cdots +
a_1 x + a_0$ where the polynomial $P - \mu$ has two simple roots and
the polynomial $P + \mu$ has $2g$ simple roots $(\mu > 0)$
\ct{fuzzy-riemann}. The unperturbed surfaces in (I)-(III) correspond to
the polynomial \eq{poly-g} with $F(X) = 0$, i.e., $M =
G_{F=0}^{-1}(\{0\})$ where $\rho = (-1)^\sharp R^2$. After
determining the embedding coordinate \eq{fluctuation} by solving the
polynomial equation $G(X) = 0$ as illustrated in the simple cases
(I)-(III), the metric of the two-dimensional surface $M =
G^{-1}(\{0\})$, according to the map \eq{nc-vector}, will be given
by the vector fields
\be \la{drei-bein-gen}
V_A = \theta^{ab} \frac{\partial X_{A}(y)}{\partial
y^b}\frac{\partial }{\partial y^a}.
\ee
The resulting metric $ds^2 = \mathfrak{G}_{ab}(y) dy^a dy^b$ where
\be \la{2-emetric-gen}
\mathfrak{G}_{ab} =
g_{AB} \frac{\partial X^{A}}{\partial y^a} \frac{\partial
X^{B}}{\partial y^b}
\ee
will again be equivalent to the induced metric on $M$ embedded in
the three-dimensional spacetime $ds^2 = g_{AB} dX^A dX^B$ whose
embedding is defined by the polynomial
\eq{poly-g}.

If we consider a generic fluctuation described by an arbitrary
polynomial \eq{poly-g}, we expect that the perturbation
\eq{fluctuation} falls into one of the three classes; (A) metric
preserving coordinate transformations generated by flat connections,
(B) volume-preserving metric deformations, and (C) volume-changing
deformations. From the analysis in Eq.\eq{inverse-emergent-metric}
we well understand for the case (A) what is going on there. The gauge
field fluctuation in Eq.\eq{fluctuation} should belong to a pure
gauge, i.e., $F^{AB} = 0$. To check this result, consider a pure
gauge ansatz $A^A(y) = g^{-1}(y)
\{ L^A, g(y) \}_\theta$. One can calculate the corresponding field
strength \eq{poisson-field}
\be \la{field-pure}
F^{AB} = \{ A^A, A^B \}_\theta
\ee
and the Casimir invariant \eq{casimir}
\be \la{casimir-pure}
g_{AB} (X^A X^B - L^A L^B) = g_{AB} A^A A^B
\ee
where $g_{AB}L^A A^B = 0$ was used. The nonvanishing terms,
$\CO(\theta^3)$ and $\CO(\theta^2)$, in Eq.\eq{field-pure} and
Eq.\eq{casimir-pure}, respectively, can be neglected in the
commutative limit and eventually will disappear in the NC space
\eq{fuzzy-sphere} as will be shown later. The case (B) corresponds to
the metric change generated by a general vector field $X$ satisfying
$\CL_X B + dA = 0$. In this case the vector field $X$ is not a
Hamiltonian vector field and it in general contains a harmonic part
in $H^1(M)$. Therefore, it could be possible that the metric
deformation generated by the nontrivial vector field $X$ will in
general accompany a topology change of the two-dimensional surface
$M$. The topology change will be triggered by a higher order, e.g.
quartic, polynomial $F(X)$ in Eq.\eq{poly-g}
\ct{fuzzy-riemann}. Finally, as a simple example of the case (C), a
volume change of the two-dimensional surface $M$ is described by the
gauge field $A^A(y) =
\alpha L^A(y)$ and $F^{AB} (y) = - \alpha (1 +
\alpha) {\varepsilon^{AB}}_C L^C =  - \alpha {\varepsilon^{AB}}_C
X^C $. In this case the Poisson bracket relation
\eq{poisson-xx} is given by
\be \la{scale-poisson}
\{ X^A, X^B \}_\theta = - (1 + \alpha) {\varepsilon^{AB}}_C X^C.
\ee
That is, the volume change can be done by turning on the polynomial
$F(X) = - \frac{\alpha}{2} g_{AB} X^A X^B$ in Eq.\eq{poly-g}.
Therefore, the volume change in Eq.\eq{casimir}, $R \to (1+\alpha)
R$, can also be interpreted as the change of coupling constant in
Eq.\eq{eom-su2}, $\lambda \to (1+\alpha)\lambda$, or the change of
noncommutativity in Eq.\eq{2-poisson}, $\theta^{ab} \to (1+\alpha)
\theta^{ab}$.

Because the Lie algebra \eq{eom-su2} arises as the equations of motion
of the action \eq{master-action}, it is necessary to generalize the
action \eq{master-action} in order to describe a general
two-dimensional surface defined by the polynomial \eq{poly-g}. The
generalized action will be defined by
\begin{equation} \la{general-action}
S_G = \Tr \Big( \frac{g^2_{YM}}{2} P_A P^A + \lambda P^A
\frac{\partial G(X)}{\partial X^A} +
\frac{i \kappa}{3!} \varepsilon_{ABC} X^A [X^B, X^C] \Big).
\ee
The equations of motion are now given by
\be \la{eom-general}
P_A = - \kappa \frac{\partial G(X)}{\partial X^A}, \qquad - \Big[
P^B \frac{\partial^2 G(X)}{\partial X^A \partial X^B} \Big] =
\frac{i}{2 g_{YM}^2} \varepsilon_{ABC} [X^B, X^C]
\ee
where $\Big[ P^B \frac{\partial^2 G(X)}{\partial X^A \partial X^B}
\Big]$ is a formal expression of the matrix ordering under the trace
for the variation $P^B \frac{\delta}{\delta X^A} \Big(
\frac{\partial G}{\partial X^B} \Big)$.
The previous equations of motion, \eq{eom-p} and \eq{eom-x}, are
given by the polynomial \eq{poly-g} with $F(X) = 0$. Of course a
vacuum manifold defined by the new action \eq{general-action} should
be newly determined by solving the equations of motion
\eq{eom-general}.

A two-dimensional NC space can be obtained by quantizing the
symplectic manifold $(M, \omega = - dy^1 \wedge dy^2)$, i.e., by
replacing the Poisson bracket \eq{2-poisson} by a star commutator
\be \la{quantization}
\{f, g\}_\theta \to -i [\widehat{f}, \widehat{g} ]_\star
\ee
and the ordinary product in $C^\infty (M)$ by the star product in NC
$\star$-algebra $\CA_\theta$. Then the local Darboux coordinates
$y^a \; (a=1,2)$ satisfy the commutation relation
\be \la{fuzzy-sphere}
[y^a, y^b]_\star = i \theta^{ab}.
\ee
The fluctuation in Eq.\eq{fluctuation} now becomes an element in
$\CA_\theta$ given by
\be \la{fuzzy-matrix}
\widehat{X}^A (y) = \widehat{L}^A(y) + \widehat{A}^A(y)
\ee
where $\widehat{L}^A(y)$ is a background solution satisfying the
constraint $(-1)^\sharp R^2 = g_{AB} \widehat{L}^A \star
\widehat{L}^B$ and $[\widehat{L}^A , \widehat{L}^B]_\star = - i
{\varepsilon^{AB}}_C \widehat{L}^C$ obtained from
Eq.\eq{lie-poisson} by the quantization \eq{quantization}. (See
\ct{fuzzy-ads} for the deformation quantization of hyperbolic planes.)
Then one can calculate the star commutator
\bea \la{fuzzy-xx}
[\widehat{X}^A, \widehat{X}^B]_\star & = & [ \widehat{L}^A(y) +
\widehat{A}^A(y), \widehat{L}^B(y) + \widehat{A}^B(y)]_\star \xx
&=& - i {\varepsilon^{AB}}_C \widehat{X}^C + [\widehat{L}^A,
\widehat{A}^B]_\star - [\widehat{L}^B,
\widehat{A}^A]_\star + [\widehat{A}^A, \widehat{A}^B]_\star
+ i {\varepsilon^{AB}}_C \widehat{A}^C \xx & = & - i
{\varepsilon^{AB}}_C \widehat{X}^C(y) + i \widehat{F}^{AB}(y).
\eea
Substituting the above expression into the action \eq{mass-cs} leads
to the action for the fluctuations
\be \la{fluc-2action}
\widehat{S}_{mCS} = - \frac{\kappa}{12 \pi |\theta|} \int d^2 y \Big(
\varepsilon_{ABC} \widehat{X}^A \star \widehat{F}^{BC} +
\lambda \widehat{X}_A \star \widehat{X}^A \Big).
\ee
The equations of motion derived from the variation with respect to
$\widehat{A}^A$ say that the fluctuations should be a flat
connection, i.e., $\widehat{F}^{AB} = 0$, already inferred from
Eq.\eq{fuzzy-xx}.

In order to treat the generalized action \eq{general-action}, the
Jacobi identity, $\varepsilon_{ABC}[\widehat{X}^A, [\widehat{X}^B,
\widehat{X}^C]_\star]_\star = 0$, can be solved in a similar way as
the commutative case by the form
\be \la{potential-field-nc}
\widehat{F}^{AB}(\widehat{X}) = \varepsilon^{ABC}
\frac{\partial \widehat{F}(\widehat{X})}{\partial \widehat{X}^C}.
\ee
The derivative $\frac{\partial \widehat{F}(\widehat{X})}{\partial
\widehat{X}^C}$ will be defined with the symmetric Weyl ordering
\ct{fuzzy-riemann}. Then one can evaluate the commutator $
[\widehat{X}^A, \frac{\partial \widehat{F}(\widehat{X})}{\partial
\widehat{X}^A}]_\star$ by a successive application of the Leibniz rule
$[\widehat{X}^A, \widehat{f} \star \widehat{g}]_\star =
\widehat{f} \star [\widehat{X}^A,  \widehat{g}]_\star +
[\widehat{X}^A, \widehat{f} ]_\star \star \widehat{g}$ such that
each term finally has a form $\widehat{F}_1 (\widehat{X}) \star
[\widehat{X}^A, \widehat{X}^B ]_\star \star \widehat{F}_2
(\widehat{X})$. If we formally denote the resulting expression as
the form
\be \la{jacobi-nc}
\half \varepsilon_{ABC}[\widehat{X}^A, [\widehat{X}^B, \widehat{X}^C]_\star]_\star
= i [\widehat{X}^A, \frac{\partial \widehat{F}
(\widehat{X})}{\partial \widehat{X}^A }]_\star = i
\Big \{ \frac{\partial^2 \widehat{F} (\widehat{X})}{\partial \widehat{X}^A \partial \widehat{X}^B}
\star [\widehat{X}^A, \widehat{X}^B ]_\star \Big\},
\ee\
it turns out that the polynomial $\frac{\partial^2 \widehat{F}
(\widehat{X})}{\partial \widehat{X}^A \partial \widehat{X}^B}$ is
symmetric with respect to $(A \leftrightarrow B)$ and so
Eq.\eq{jacobi-nc} identically vanishes. Therefore, the star
commutator \eq{fuzzy-xx} takes the form \ct{fuzzy-riemann}
\be \la{comm-potential}
[\widehat{X}^A, \widehat{X}^B]_\star = - i \varepsilon^{ABC}
\frac{\partial \widehat{G}(\widehat{X})}{\partial \widehat{X}^C}
\ee
where the polynomial $\widehat{G}(\widehat{X})$ is the star product
version of Eq.\eq{poly-g} given by
\be \la{poly-g-nc}
\widehat{G}(\widehat{X}) = \widehat{F}(\widehat{X})
- \half g_{AB} \widehat{X}^A \star \widehat{X}^B + \rho.
\ee

Suppose that we have solved the polynomial equation
$\widehat{G}(\widehat{X}) = 0$ whose solution is given by
$\widehat{X}_A = g_{AB}\widehat{X}^B = \widehat{X}_A(y)$. (See
\ct{fuzzy-riemann} for explicit solutions for tori and deformed spheres.)
Now one can define an inner derivation of the NC $\star$-algebra
$(\CA_\theta, [\cdot, \cdot]_\star)$ as in Eq.\eq{nc-vector} by
considering an adjoint action of $\widehat{X}_A(y) =
g_{AB}\widehat{X}^B(y)$ as follows
\bea \la{3-vector}
\widehat{V}_A [\widehat{f}] (y) &\equiv & ad_{\widehat{X}_A} [\widehat{f}] (y)
= - i [ \widehat{X}_A(y), \widehat{f}(y) ]_\star \xx & = & V_A^a (y)
\frac{\partial f(y)}{\partial y^a} + {\cal O}(\theta^3).
\eea
The leading term in Eq.\eq{3-vector} is exactly equal to the vector
fields $V_A (y) = V_A^a (y) \frac{\partial }{\partial y^a}$ in
Eq.\eq{drei-bein-gen}. We may identify $\widehat{V}_A$ with
generalized tangent vectors defined on a two-dimensional fuzzy
manifold described by the polynomial \eq{poly-g-nc}.

As was shown in Eq.\eq{semi-gauge-tr}, the symplectomorphism can be
identified with NC $U(1)$ gauge transformations. Flat connections,
i.e., $\widehat{F}^{AB}(y) = 0$ in which case
$\widehat{F}(\widehat{X}) = 0$, are given by $\widehat{A}^A(y) =
\widehat{g}^{-1}(y) \star [\widehat{L}^A(y),
\widehat{g}(y)]_\star$ or $\widehat{X}^A(y) = \widehat{g}^{-1}(y)
\star \widehat{L}^A(y) \star \widehat{g}(y)$ with any invertible $\widehat{g}(y)
\in \CA_\theta$. So the equations of motion \eq{fuzzy-xx} are the same as before
and the solution \eq{fuzzy-matrix} of flat connections preserves the
area \eq{casimir}, say, $g_{AB} \widehat{X}^A \star \widehat{X}^B =
(-)^\sharp R^2$. Also note that the remaining terms in
Eqs.\eq{field-pure} and \eq{casimir-pure} are completely cured in
the NC space \eq{fuzzy-sphere} as we remarked before. Because the
embedding \eq{poly-g-nc} has not been changed, it is a natural
consequence that a pure gauge fluctuation does not change a
two-dimensional metric of fuzzy manifold $\widehat{M}$ as we already
noted in Eq.\eq{inverse-emergent-metric}.

Now we want to discuss some interesting aspects of our construction.
As we observed above, a pure gauge fluctuation does not change the
two-dimensional metric $ds^2 = \mathfrak{G}_{ab} dy^a dy^b$ and
belongs to the same representation, i.e., $g_{AB} L^A L^B = g_{AB}
X^A X^B = (-)^\sharp R^2$ for $L^A(y)$ and $X^A(y) = L^A(y) +
A^A(y)$ in $C^\infty(M)$. This means that there exists a global
Lorentz transformation in three dimensions such that $X^A =
{\Lambda^A}_B L^B$ where ${\Lambda^A}_B
\in SO(3-p, p)$. In other words the metric
$\mathfrak{G}_{ab}$ is invariant under the Lorentz transformation in
ambient spaces as expected. It is interesting to notice that a local
gauge transformation in two dimensions can be interpreted as a
global Lorentz transformation in three-dimensional target spacetime.
More generally, one may represent a generic fluctuation of gauge
fields in $X^A$ as a general coordinate transformation, that is,
$L^A(y) \mapsto X^A(y) = X^A(L)(y)$. Then the vector fields $V_A^L$
and $V_A^X$ in $TM$ for the smooth functions $L^A(y)$ and $X^A(y)$
are defined by Eq.\eq{drei-bein-gen} and they are related by $V_A^X
= \frac{\partial X_A}{\partial L_B}(y) V_B^L$ thanks to the chain
rule $\{X_A(L), f \}_\theta = \frac{\partial X_A}{\partial L_B}
\{L_B , f \}_\theta$. According to Eq.\eq{2-emetric-gen}, the
two-dimensional metric can then be written as
\be \la{co-metric}
\mathfrak{G}_{ab} = G_{AB} \frac{\partial L^A}{\partial y^a}
\frac{\partial L^B}{\partial y^b}
\ee
where $G_{AB}(y) = \frac{\partial X^C}{\partial L^A}(y)
\frac{\partial X^D}{\partial L^B}(y) g_{CD}$. Thus a generic fluctuation possibly
changing the volume as well as topology [turning on a nontrivial
$F(X) \neq 0$] can be interpreted as a general coordinate
transformation supported on the two-dimensional surface $M$. Of
course this is consistent with the fact that the metric $ds^2 =
\mathfrak{G}_{ab} dy^a dy^b$ is the induced metric on a submanifold
$M$ embedded in $\IR^{3-p, p}$.

It is well known \ct{deser-jackiw} that the massive Chern-Simons
gauge theory in three dimensions has a physical degree of freedom.
One may wonder which mode in the action \eq{mass-cs} corresponds to
the physical one. Note that the gauge field dynamics in three
dimensions need not be subject to the constraint \eq{casimir}. Because
gauge field fluctuations preserving a two-dimensional area and
satisfying the equations of motion \eq{eom-su2} are flat connections
and also pure gauges, the only remaining physical mode satisfying the
same Lie algebra \eq{eom-su2} is an area changing fluctuation as we
observed in Eq.\eq{scale-poisson}. Because the area change can also be
interpreted as the change of coupling constant or noncommutativity,
it would be intriguing to recall that a similar feature also arises
in the AdS/CFT correspondence
\ct{ads-cft} where the size of bulk spacetime is related to the coupling
constant of gauge theory.

\section{Emergent Geometry for Snyder Spacetime}

Now we want to generalize the analysis for the two-dimensional cases
to higher dimensions, in particular, to four-dimensional manifolds
with constant curvature as an emergent geometry from some matrix
model. Let us start with the following IKKT matrix model with a mass
deformation \ct{mass-matrix}:
\be \la{m-ikkt}
S_{mIKKT} = \Tr \big( -\frac{1}{4} [X^a, X^b]^2 + \frac{(d-1)\kappa}{2}
X_a X^a \big)
\ee
where $X^a$ are $N \times N$ Hermitian matrices and $a,b=1, \cdots, d \geq 2$.
One can rewrite the action \eq{m-ikkt} as the form
\be \la{mass-ikkt}
S_\kappa = \Tr \big( \frac{1}{4} M_{ab} M^{ab} -
\frac{1}{2} M_{ab} [X^a, X^b] + \frac{(d-1)\kappa}{2}
X_a X^a \big)
\ee
by introducing Lagrange multipliers $M_{ab}$ which are $N \times N$ anti-Hermitian matrices.
In spite of the mass deformation with $\kappa \neq 0$, the matrix action
\eq{mass-ikkt} respects the $U(N)$ gauge symmetry given by
\be \la{un-gauge-symm}
(X^a, M_{ab}) \to  U (X^a, M_{ab}) U^\dagger
\ee
with $U \in U(N)$. The equations of motion are given by
\bea \la{eom1-matrix}
&& [X^a, X^b] = M^{ab}, \\
\la{eom2-matrix}
&& [M^{ab}, X_b] + (d-1) \kappa  X^a = 0,
\eea
where Eq.\eq{eom2-matrix} becomes the equations of motion derived
from the action \eq{m-ikkt} when substituting $[X^a,X^b]$
for $M^{ab}$. One can easily check that the above equations of motion can be
obtained from the Snyder algebra \ct{snyder}:
\bea \la{snyder-alg}
&& [X^a, X^b] = M^{ab}, \xx && [X^a, M^{bc}] =  \kappa \Big( g^{ac}
X^b - g^{ab} X^c \Big), \\
&& [M^{ab}, M^{cd}] =  \kappa \Big(g^{ac} M^{bd} - g^{ad} M^{bc} - g^{bc} M^{ad}
+ g^{bd} M^{ac} \Big), \nonumber
\eea
where the last equation can be derived from the other two applying
the Jacobi identity. Therefore, if the matrices $(X^a, M^{bc})$
satisfy the Snyder algebra \eq{snyder-alg}, they automatically
satisfy the equations of motion, \eq{eom1-matrix} and
\eq{eom2-matrix}. Here the deformation parameter $\kappa$ carries the physical dimension of
$({\rm length})^2$ since we will consider $X^a$ as ``matrix coordinates."

Because we consider the action \eq{mass-ikkt} as a massive
deformation of the IKKT matrix model \eq{ikkt}, we regard the matrices
$M_{ab}$ in the action \eq{mass-ikkt} as Lagrange multipliers
and so these can be integrated out. The resulting action of course recovers
the original action \eq{m-ikkt}. Thus the number of dynamical coordinates remains the same as
the undeformed case. Actually it will be shown later that
the matrix $X^a$ as a dynamical coordinate is mapped to a NC gauge field
and $M^{ab}$ to its field strength. Therefore, the emergent geometry
for the mass-deformed case can be derived by essentially the same way as the
undeformed case, except that the deformed case in general admits a
Poisson structure only instead of a symplectic structure. But this
is not a difficulty since a Poisson structure is enough to formulate
emergent geometry from large $N$ matrices or NC gauge fields, as will be shown below.
Note that Poisson manifolds are a more general class of manifolds which contains
symplectic manifolds as a special class.

Now the problem is how to generalize the emergent geometry picture
for the undeformed case \eq{ikkt} to the mass-deformed case
\eq{mass-ikkt} where the vacuum geometry will be nontrivial, i.e.,
curved, since $M^{ab} =$ constants cannot be a vacuum solution
unlike the $\kappa = 0$ case. We showed that the generators $X^a$ in the Snyder
algebra \eq{snyder-alg} satisfy the equations of motion
\eq{eom1-matrix} and \eq{eom2-matrix}. In order to map the matrix algebra
$(M_N, [\cdot,\cdot])$ defining the action \eq{mass-ikkt} to a NC
$\star$-algebra $(\CA_\theta, [\cdot,\cdot]_\star)$, we will show
that the Snyder algebra \eq{snyder-alg} can be obtained by the
deformation quantization of a Poisson manifold \ct{kontsevich} whose
Poisson tensor is given by $\Pi = \half L^{ab}(x)
\frac{\partial}{\partial x^a} \wedge
\frac{\partial}{\partial x^b}$. In other words, we want to show that
the Schouten bracket \ct{poisson-book} for the Poisson tensor $\Pi$
vanishes, i.e.,
\be \la{schouten}
 [\Pi, \Pi]_S \equiv \Big( L^{da} \frac{\partial L^{bc}}{\partial x^d} +
L^{db} \frac{\partial L^{ca}}{\partial x^d} + L^{dc} \frac{\partial
L^{ab}}{\partial x^d} \Big) \frac{\partial}{\partial x^a}
\wedge \frac{\partial}{\partial x^b} \wedge \frac{\partial}{\partial
x^c} = 0
\ee
if the Poisson bracket $\{ x^a, x^b \}_\Pi = L^{ab}(x) = \langle
\Pi, dx^a \wedge dx^b \rangle$ satisfies the Snyder algebra \eq{snyder-alg}.
It is easy to see that the Jacobi identity $\{ \{ x^a, x^b \}_\Pi,
x^c \}_\Pi + \{ \{ x^b, x^c \}_\Pi, x^a \}_\Pi + \{ \{ x^c, x^a
\}_\Pi, x^b \}_\Pi = 0$ is satisfied due to the second algebra in
Eq.\eq{snyder-alg}. From the Jacobi identity, we immediately get the
result \eq{schouten} and so the two-vector field $\Pi$ is a Poisson
tensor.

The Poisson tensor $\Pi$ of a Poisson manifold $M$ induces a bundle
map $\Pi^\sharp: T^* M \to TM$ by
\be \la{anchor}
A \mapsto \Pi^\sharp(A) = L^{ab}(x) A_a(x)
\frac{\partial}{\partial x^b}
\ee
for $A = A_a(x) dx^a \in T_x^* M$, which is called the anchor map of
$\Pi$ \ct{poisson-book}. The rank of the Poisson structure at a
point $x \in M$ is defined as the rank of the anchor map at this
point. If the rank equals the dimension of the manifold at each
point, the Poisson structure reduces to a symplectic structure which
is also called nondegenerate. The nondegenerate Poisson structure
uniquely determines the symplectic structure defined by the two-form
$\omega =
\half \omega_{ab}(x) d x^a \wedge d x^b = \Pi^{-1}$ and the
condition \eq{schouten} is equivalent to the statement that the
two-form $\omega$ is closed, $d\omega = 0$. In this case the anchor
map $\Pi^\sharp: T^* M \to TM$ is a bundle isomorphism as we
discussed in Sec. 1. To define a Hamiltonian vector field
$\Pi^\sharp(df)$ of a smooth function $f \in C^\infty(M)$, what one
really needs is a Poisson structure which reduces to a symplectic
structure for the nondegenerate case. Given a smooth Poisson
manifold $(M, \Pi)$, the map $f \mapsto X_f = \Pi^\sharp(df)$ is a
homomorphism \ct{poisson-book} from the Lie algebra $C^\infty(M)$ of
smooth functions under the Poisson bracket to the Lie algebra of
smooth vector fields under the Lie bracket. In other words, the Lie
algebra homomorphism \eq{poisson-lie} is still true even for any
Poisson manifold.

Like the Darboux theorem in symplectic manifolds, the Poisson
geometry also enjoys a similar property known as the splitting
theorem proved by Weinstein \ct{weinstein}. The splitting theorem
states that a $d$-dimensional Poisson manifold is locally equivalent
to the product of $\IR^{2n}$ equipped with the canonical symplectic
structure with $\IR^{d-2n}$ equipped with a Poisson structure of
rank zero at the origin. That is, the Poisson manifold $(M, \Pi)$ is
locally isomorphic (in a neighborhood of $x$) to the direct product
$S \times N$ of a symplectic manifold $(S, \sum_{i=1}^n dq^i \wedge
dp_i)$ with a Poisson manifold $(N_x, \{\cdot, \cdot\}_N)$ whose
Poisson tensor vanishes at $x$.

Note that not every Snyder space can be obtained by the quantization
of a symplectic manifold $(M, \omega)$ in contrast to the
two-dimensional orientable  hyperspaces in Sec. 2. If $M$ is
a compact symplectic manifold, the second de Rham cohomology group
$H^2(M)$ is nontrivial and so the only $n$-sphere that admits a
symplectic form is the two-sphere. For example, let ${\bf S}^4 =
\{(u, v, t) \in \IC \times \IC \times \IR: |u|^2 + |v|^2 =
t(2-t) \}$. Then the bivector field $\Pi = u v
\partial_u \wedge \partial_v - u v^* \partial_u \wedge
\partial_{v^*} - u^* v \partial_{u^*} \wedge \partial_v
+ u^* v^* \partial_{u^*} \wedge \partial_{v^*}$ is a Poisson tensor,
that is, $[\Pi, \Pi]_S =0$, and $\Pi \wedge \Pi = 4 |u|^2 |v|^2
\partial_u \wedge \partial_{v} \wedge
\partial_{u^*} \wedge \partial_{v^*}$.
Therefore, the Poisson tensor $\Pi$ vanishes on a subspace of either
$u = 0$ or $v =0$ and the Poisson structure becomes degenerate
there. This is the reason why we have to rely on a Poisson structure
rather than a symplectic structure to formulate emergent geometry
from the Snyder algebra \eq{snyder-alg}.

Because any Poisson manifold can be quantized via deformation
quantization \ct{kontsevich}, the anchor map \eq{anchor} can be
lifted to a NC manifold as in Eq.\eq{nc-vector}. As we noticed
before, it is enough to have a Poisson structure to achieve the map
$C^\infty(M) \to \Gamma(TM): f \mapsto X_f = \Pi^\sharp(df)$ such as
Eq.\eq{duality}. So let us take the limit $N \to
\infty$ of the Snyder algebra \eq{snyder-alg} and suppose that the
Poisson manifold $(M, \Pi)$ is quantized via deformation
quantization, i.e.,
\be \la{snyder-space}
\{ x^a, x^b \}_\Pi = L^{ab}(x) \;\; \to \;\;
[\widehat{x}^a, \widehat{x}^b]_{\star} = i
\widehat{L}^{ab}(\widehat{x})
\ee
where $\widehat{L}^{ab}(\widehat{x}) \in \CA_\theta$ are assumed to be
dimensionful operators of $(\rm{length})^2$ satisfying the Snyder
algebra \eq{snyder-alg}.

Let us consider a vacuum solution of the mass-deformed matrix model
\eq{mass-ikkt} as the Snyder space defined by Eq.\eq{snyder-space}.
Now we will regard the background solution $\widehat{x}^a \in \CA_\theta$
in \eq{snyder-space} as NC fields but from now on we will omit
the hat for notational simplicity. Consider fluctuations of
the large $N$ matrices $X^a \equiv \kappa g^{ab} \widehat{D}_b(x)$
around the vacuum solution \eq{snyder-space} as follows
\be \la{d-exp}
\widehat{D}_a(x) = \widehat{D}^{(0)}_a(x) + \widehat{A}_a(x)
\ee
where $\widehat{D}^{(0)}_a(x) = \frac{1}{\kappa} g_{ab} x^b$. The
background solution $(\kappa \widehat{D}^{(0)}_a(x), \widehat
{L}_{ab}(x) \equiv \kappa^2 \widehat{Q}_{ab}
)$ satisfies the Snyder algebra
\eq{snyder-alg}. Using the above variables, one can calculate the
star commutator
\bea \la{comm-xx}
[\widehat{D}_a, \widehat{D}_b]_\star & = & [ \widehat{D}^{(0)}_a(x)
+ \widehat{A}_a(x), \widehat{D}^{(0)}_b (x) +
\widehat{A}_b(x)]_\star \xx
&=& i \widehat{Q}_{ab} + [ \widehat{D}^{(0)}_a,
\widehat{A}_b]_\star - [ \widehat{D}^{(0)}_b, \widehat{A}_a]_\star
+ [\widehat{A}_a, \widehat{A}_b]_\star
\xx & \equiv & i \widehat{F}_{ab}.
\eea
One can check that the field strength defined in Eq.\eq{comm-xx}
covariantly transforms under the gauge transformation $\delta
\widehat{A}_a = -i \big( [\widehat{D}^{(0)}_a, \widehat{\lambda}]_\star
+ [\widehat{A}_a, \widehat{\lambda}]_\star \big)$, viz.,
\be \la{cov-tr-f}
\delta \widehat{F}_{ab} = -i [ \widehat{F}_{ab}, \widehat{\lambda}]_\star.
\ee
Note that we need the background part $\widehat{Q}_{ab}$ in
$\widehat{F}_{ab}$ to maintain the gauge covariance \eq{cov-tr-f}.
Using the result \eq{comm-xx}, we get the action for the
fluctuations after integrating out the $M$-fields in
Eq.\eq{mass-ikkt}
\be \la{mass-ncu1}
\widehat{S}_{\kappa} =  \frac{\kappa^4}{4} \Tr_{\CH} g^{ac} g^{bd}
\widehat{F}_{ab} \star \widehat{F}_{cd} + \frac{(d-1) \kappa^3}{2}
\Tr_{\CH}  g^{ab} \widehat{D}_a  \star \widehat{D}_b
\ee
where the trace $\Tr_{\CH}$ is defined over the Hilbert space $\CH$
associated with a representation space of the NC $\star$-algebra
\eq{snyder-space}. It might be remarked that, in spite of the mass
term, the action \eq{mass-ncu1} respects the NC $U(1)$ gauge
symmetry acting on $(\widehat{D}_a,
\widehat{F}_{ab})
\to \widehat{U} \star (\widehat{D}_a, \widehat{F}_{ab}) \star
\widehat{U}^\dagger$ where $\widehat{U} \in \CA_\theta$.

Because $\widehat{D}_a = \frac{1}{\kappa} g_{ab} X^b$, one can rewrite
the Snyder algebra \eq{snyder-alg} in terms of gauge theory
variables:
\bea \la{snyder-gauge}
&& [\widehat{D}_a, \widehat{D}_b]_\star = i \widehat{F}_{ab} =
\kappa^{-2} M_{ab}, \xx && [\widehat{D}_a, \widehat{F}_{bc}]_\star = - i
\kappa^{-1} (g_{ac} \widehat{D}_b - g_{ab} \widehat{D}_c), \\
&& [\widehat{F}_{ab}, \widehat{F}_{cd}]_\star = - i \kappa^{-1}
(g_{ac} \widehat{F}_{bd} - g_{ad} \widehat{F}_{bc} - g_{bc}
\widehat{F}_{ad} + g_{bd} \widehat{F}_{ac}). \nonumber
\eea
Because $-i [\widehat{D}_a, \widehat{F}_{bc}]_\star = -i [x_a/\kappa,
\widehat{F}_{bc}]_\star - i [\widehat{A}_a, \widehat{F}_{bc}]_\star \equiv \widehat{D}_a
\widehat{F}_{bc}$, one can easily check that the Bianchi identity,
$\widehat{D}_{[a} \widehat{F}_{bc]} = 0$, and the equations of
motion, $\widehat{D}_a
\widehat{F}^{ab} = (d-1)
\kappa^{-1} \widehat{D}^b$, are directly derived from the second algebra
in Eq.\eq{snyder-gauge}. Note that the last equation in
Eq.\eq{snyder-gauge} can be obtained from the other two applying the
Jacobi identity. From a gauge theory point of view, it is a bizarre
relation since the field strength $\widehat{F}_{ab}$ of an arbitrary
gauge field $\widehat{A}_a$ behaves like an angular momentum
operator in $d$-dimensions. This kind of behavior is absent in
an undeformed case, $\kappa = 0$. The theory will strongly constrain
the behavior of gauge fields and so there might be some hidden
integrability.

A Hamiltonian vector field $X_f = \Pi^\sharp(df)$ for a smooth
function $f \in C^\infty(M)$ is defined by the anchor map
\eq{anchor} as follows \ct{poisson-book}:
\be \la{ham-vec}
X_f(g) = - \langle \Pi, df \wedge dg \rangle = - L^{ab}(x)
\frac{\partial f}{\partial x^a} \frac{\partial g}{\partial x^b} =
\{g, f\}_\Pi.
\ee
Because the Poisson manifold $(M, \Pi)$ has been quantized in
Eq.\eq{snyder-space}, the correspondence between the Lie algebras
$(C^\infty(M), \{\cdot, \cdot \}_\Pi)$ and $(\Gamma(TM), [\cdot,
\cdot])$ can be lifted to the NC $\star$-algebra $(\CA_\theta, [\cdot,
\cdot]_\star)$ as in Eq.\eq{nc-vector}. That is, we can map NC fields
in $\CA_\theta$ to vector fields in $\Gamma_\theta (\widehat{TM})$,
$\CA_\theta$-valued sections of a generalized tangent bundle
$\widehat{TM}$. For example, $\widehat{D}_a(x)$ in Eq.\eq{d-exp} are
mapped to the following vector fields in $\widehat{TM}$
\bea \la{mass-vector}
ad_{\widehat{D}_a} [\widehat{f}] (x) &\equiv & - i [
\widehat{D}_a(x), \widehat{f}(x) ]_{\star}
= - L^{\mu\nu}(x)  \frac{\partial D_a(x)}{\partial x^\nu}
\frac{\partial f(x)}{\partial x^\mu} +
\cdots \xx &\equiv & V^\mu_a (x) \frac{\partial f(x)}{\partial x^\mu}
+ \cdots = V_a[f](x) + {\cal O}(L^3)
\eea
where the leading order leads to the usual vector fields $V_a
\in TM$ in Eq.\eq{ham-vec}.

We might express from the outset the star product using different NC
coordinates $\widehat{y}^a$ defined by $[\widehat{y}^a,
\widehat{y}^b]_{\widetilde{\star}} = i \widehat{\widetilde{L}}^{ab}(\widehat{y})$.
In terms of the new $\widetilde{\star}$-product, the adjoint action
defining an inner derivation in $\CA_\theta$ is then given by
\bea \la{new-nc-der}
ad_{\widehat{D}_a} [\widehat{f}] (\widehat{y}) &\equiv & - i [
\widehat{D}_a(\widehat{y}), \widehat{f}(\widehat{y}) ]_{\widetilde{\star}}
= - \widetilde{L}^{\mu\nu}(y)  \frac{\partial
D_a(y)}{\partial y^\nu} \frac{\partial f(y)}{\partial y^\mu} +
\cdots \xx &\equiv & \widetilde{V}^\mu_a (y) \frac{\partial f(y)}{\partial y^\mu}
+ \cdots = \widetilde{V}_a[f](y) + {\cal O}(\widetilde{L}^3).
\eea
Noting that the star products, $\star$ and $\widetilde{\star}$, are
related by a coordinate transformation $x^a \mapsto y^a =
y^a(x)$ \ct{kontsevich}, in other words,
$$
\widetilde{L}^{\mu\nu}(y) = L^{ab}(x) \frac{\partial
y^\mu(x)}{\partial x^a}\frac{\partial y^\nu(x)}{\partial x^b},
$$
one can easily check \ct{hsy1} using the chain rule that the vector
fields defined by Eq.\eq{new-nc-der} are diffeomorphic to those in
Eq.\eq{mass-vector} as expected, i.e.,
\be \la{diff-vec}
\widetilde{V}_a^\mu (y) = V_a^\nu (x) \frac{\partial y^\mu(x)}{\partial
x^\nu}.
\ee

We are particularly interested in the background geometry defined by
Eq.\eq{snyder-space}. In this case, the vector fields for the background
gauge fields $\widehat{D}^{(0)}_a(x)$ are given by
\bea \la{nc-yvector}
ad_{\widehat{D}^{(0)}_a} [\widehat{f}] (x) & = & - i [
\widehat{D}^{(0)}_a (x), \widehat{f}(x) ]_\star = V_a^{(0)\mu} (x) \frac{\partial f(x)}{\partial
x^\mu} + \cdots \xx &\equiv & V^{(0)}_a[f](x) + {\cal O}(L^3).
\eea
Using the Snyder algebra for $(\widehat{D}^{(0)}_a,
\widehat{Q}_{ab})$ and the relation
\bea \la{vec-bianchi}
&& ad_{[\widehat{D}^{(0)}_a, \widehat{D}^{(0)}_b]_\star} = i
[V^{(0)}_a, V^{(0)}_b] + {\cal O}(L^3) \xx && = i \;
ad_{\widehat{Q}_{ab}}
\equiv  i S^{(0)}_{ab} + {\cal O}(L^3),
\eea
one can see that $(V^{(0)}_a, S^{(0)}_{ab}) \in \Gamma(TM_{{\rm
back}})$ also satisfy the Snyder algebra \eq{snyder-alg} where the
Lie algebra in $\Gamma(TM_{{\rm back}})$ is defined by the Lie
bracket, e.g., $[V^{(0)}_a, V^{(0)}_b] = S^{(0)}_{ab}$.

We want to find the representation of the Snyder algebra
\eq{snyder-alg} in terms of differential operators \cite{snyder}, i.e., vector fields in
$\Gamma(TM_{{\rm back}})$. In order to find an explicit expression
of vector fields $V^{(0)}_a \in \Gamma(TM_{{\rm back}})$, first
notice that the Snyder algebra
\eq{snyder-alg} can be understood as the Lorentz algebra in
$(d+1)$ dimensions with the identification $M^{d+1, a} =  \sqrt{\kappa} X^a$
\be \la{high-lorentz}
[M^{AB}, M^{CD}] = \kappa \Big( g^{AC} M^{BD} - g^{AD} M^{BC} - g^{BC} M^{AD} +
g^{BD} M^{AC} \Big)
\ee
where $A,B, \cdots = 1, \cdots, d+1$. Therefore the equivalence
between the Snyder algebra \eq{snyder-alg} in $d$ dimensions and the
Lorentz algebra $SO(d+1-p,p)$ in $(d+1)$ dimensions implies that the
Snyder space as an emergent geometry defined by the action
\eq{mass-ikkt} can be obtained as a $d$-dimensional hypersurface
$M$ embedded in $\IR^{d+1-p, p}$. For example, in the $d=2$ case,
the Lorentz algebra \eq{high-lorentz} is equivalent to the Lie
algebra \eq{eom-su2} with the identification $M^{AB} =
- i \lambda {\varepsilon^{AB}}_C X^C, \; (A,B,C =1,2,3)$ where $ \kappa
= - \lambda^2 \det g_{AB}$ and so $[X^A, X^B] = M^{AB}$. And the Lie algebra
\eq{eom-su2} describes a two-dimensional hypersurface
foliated by the quadratic form \eq{casimir} in $\IR^{3-p, p}$. See
the Appendix for the details.

Similarly we will consider, in particular, four-dimensional
hypersurfaces $M$ for three cases with $p=0,1,2$. Let us consider a
homogeneous quadratic form as an invariant of the Lorentz algebra
\eq{high-lorentz}
\be \la{de-sitter}
 g_{AB} x^A x^B = (-1)^\sharp R^2
\ee
and the ambient space metric $g_{AB}$ will be taken as a
five-dimensional flat Euclidean or Lorentzian metric given by (I)
$g_{AB} = {\rm diag}(1,1,1,1,1)$ with $\sharp = 0$, (II) $g_{AB} =
{\rm diag}(-1, 1,1,1,1)$ with $\sharp = 0$, and (III) $g_{AB} = {\rm
diag}(-1, 1,1,1,-1)$ with $\sharp = 1$. Then they describe (I) ${\bf
S}^4$, (II) $dS_4$, and (III) $AdS_4$ of radius $R$ given by
Eq.\eq{de-sitter} in a continuum limit. It should be remarked that
the case (I), a four-sphere ${\bf S}^4$, admits only a Poisson
structure instead of a nondegenerate symplectic structure \footnote{Instead one can consider a
bundle over ${\bf S}^4$ with fibre ${\bf S}^2$, which is the
K\"ahler coset space $SO(5)/U(2) \simeq {\bf S}^4 \times {\bf S}^2$
\ct{aoyama}. Then ${\bf S}^4$ may be described by the complex
coordinate system of $SO(5)/U(2)$, where a symplectic structure
manifests and the second de Rham cohomology group $H^2({\bf S}^4
\times {\bf S}^2)$ is definitely nontrivial.} and so its
quantization has to be described in terms of deformation
quantization of Poisson manifold as we explained
before. Although we do not know whether the other cases, (II) and
(III), admit a nondegenerate Poisson, i.e., symplectic, structure,
the arguments followed by Eq.\eq{high-lorentz} will be completely
sensible even with the Poisson structure only.

Suppose that $x^A = x^A(x), \; A=1,2,\cdots,5$ satisfy
Eq.\eq{de-sitter} and are local parameterizations of $M$ in terms
of local coordinates $x^a$. We will identify $x^A = x^a,
\; A = 1, \cdots, 4$, with the Poisson coordinates in Eq.\eq{snyder-space},
which is the background solution $X^a_{{\rm back}} = x^a = \kappa
g^{ab} D_b^{(0)}(x)$ in Eq.\eq{d-exp}. The vector fields
$(V^{(0)}_a, S^{(0)}_{ab}) \in
\Gamma(TM_{{\rm back}})$ in Eqs.\eq{nc-yvector} and \eq{vec-bianchi}
satisfying the Snyder algebra can then be understood as differential
Lorentz generators of $SO(5-p, p)$,
\be \la{diff-lorentz}
S^{(0)}_{AB} = \kappa \Big(x_B \frac{\partial}{\partial x^A} - x_A
\frac{\partial}{\partial x^B}\Big)
\ee
where $V^{(0)}_a \equiv S^{(0)}_{5,a}/\sqrt{\kappa}$ and the five-dimensional metric
$g_{AB}$ (to define $x_A = g_{AB} x^B$) is a standard flat metric.
According to the identification, the vector fields $V^{(0)}_a$ in
Eq.\eq{nc-yvector} can be represented by the coordinates $x^A$ as
follows \ct{snyder}
\bea \la{vec-v5}
V^{(0)}_a &=& V^{(0)A}_a (x) \frac{\partial}{\partial x^A} =
S^{(0)}_{5,a}/\sqrt{\kappa} \xx &=& \sqrt{\kappa} \Big(x_a \frac{\partial}{\partial x^5} - x_5
\frac{\partial}{\partial x^a}\Big)
\eea
and so we get the result $V^{(0)b}_a = - \sqrt{\kappa} \delta^b_a x_5$ and
$V^{(0)5}_a = \sqrt{\kappa} x_a$. Then it is obvious that $S^{(0)}_{ab} =
[V^{(0)}_a, V^{(0)}_b] = \kappa (x_b \frac{\partial}{\partial x^a} - x_a
\frac{\partial}{\partial x^b})$ are the generators of the four-dimensional Lorentz group,
i.e., $S^{(0)}_{ab} \in SO(4)$ or $SO(3,1)$ and $(V^{(0)}_a,
S^{(0)}_{ab}) $ satisfy the Snyder algebra \eq{snyder-alg}.

As will be shown in the Appendix, the Lie algebra
\eq{eom-su2} is the Snyder algebra in two dimensions whose generators
are given by $(X^1, X^2, M^{12}= \pm \frac{i}{\lambda} X^3)$.
In this case the Lie algebra \eq{eom-su2} describes a
two-dimensional hypersurface embedded in three-dimensional space
whose metric is given by Eq.\eq{2-emetric}. Therefore, in order to
define a four-dimensional metric determined by the Snyder algebra
\eq{snyder-alg}, we will consistently extend the two-dimensional
case and so the metric is defined by the vector fields \eq{vec-v5}
as follows
\bea \la{metric-snyder}
ds^2 &=& \mathfrak{G}^{(0)}_{ab} dx^a \otimes dx^b \xx &=& (\det
\mathfrak{G}^{(0)}_{ab}) g_{AB}V_a^{(0)A} V_b^{(0)B} dx^a
\otimes dx^b \xx
&=& (\det \mathfrak{G}^{(0)}_{ab}) (g_{ab} x_5^2 + g_{55} x_a x_b)
dx^a \otimes dx^b
\eea
where $\det \mathfrak{G}^{(0)}_{ab} = \frac{1}{x_5^2}$ and we put
$R= \kappa =1$ for simplicity. Of course Eq.\eq{metric-snyder} describes a
four-dimensional maximally symmetric space with a constant
curvature, e.g., ${\bf S}^4, \; dS_4$ or $AdS_4$ depending on the
signature of the five-dimensional metric $g_{AB}$. Because $x^5 =
\pm \sqrt{1 - g_{55} g_{ab} x^a x^b}$, the metric \eq{metric-snyder} can be rewritten as
the following form
\bea \la{ind-snyder}
ds^2 &=& \Big(g_{ab} + g_{55} \frac{x_a x_b}{x_5^2} \Big) dx^a
\otimes dx^b \xx &=& g_{AB} \frac{\partial x^A}{\partial x^a}
\frac{\partial x^B}{\partial x^b} dx^a \otimes dx^b \xx &=&
g_{AB} dx^A \otimes dx^B.
\eea
Note that the final result \eq{ind-snyder} is completely parallel to
the two-dimensional one, e.g., \eq{2-emetric}.

Therefore, we get an interesting result. The mass-deformed IKKT
matrix model \eq{mass-ikkt} in $d$ dimensions is completely
described by the Snyder algebra \eq{snyder-alg} which is equivalent
to the Lorentz algebra $SO(d+1-p,p)$, Eq.\eq{high-lorentz}, in
$(d+1)$ dimensions. We found that a vacuum geometry of the Snyder
algebra is a constant curvature space. For example, the metric
\eq{metric-snyder} in four dimensions describes ${\bf S}^4, \, dS_4$,
and $AdS_4$ depending on the choice of the five-dimensional metric
$g_{AB}$. Thus the equivalence between the Snyder algebra
\eq{snyder-alg} in $d$ dimensions and the Lorentz algebra
\eq{high-lorentz} in $(d+1)$ dimensions is beautifully realized as a
well-known geometrical result that a constant curvature space in
$d$ dimensions such as ${\bf S}^d, \; dS_d$, and $AdS_d$ can be
embedded in a flat Euclidean or Lorentzian spacetime in
$(d+1)$ dimensions. In particular, this result clearly illustrates
how a nontrivial curved spacetime emerges from the zero-dimensional
(i.e., background independent) matrix model \eq{mass-ikkt} through
the correspondence \eq{nc-vector} between NC $\star$-algebra
$(\CA_\theta, [\cdot, \cdot]_\star)$ and
$\Gamma_\theta(\widehat{TM})$, generalized vector fields.  We will
discuss in Sec. 4 how the constant curvature spacetimes in
Eq.\eq{ind-snyder} can be described by the coset space realization
of the Snyder algebra \eq{snyder-alg}.

We can further deduce consistent pictures about emergent geometry by
closely following the two-dimensional case we observed in the
previous section. Consider a generic fluctuation in Eq.\eq{d-exp}.
If the fluctuation is a flat connection, i.e., $\widehat{A}_a(x) =
\widehat{g}^{-1}(x) \star [\widehat{D}_a^{(0)},
\widehat{g}(x)]_\star $, then $\widehat{D}_a(x) = \widehat{g}^{-1}(x) \star
\widehat{D}_a^{(0)} \star \widehat{g}(x)$ and $\widehat{F}_{ab}(x) =
\widehat{g}^{-1}(x) \star \widehat{Q}_{ab} \star \widehat{g}(x)$.
One can immediately see from Eq.\eq{snyder-gauge} that the Snyder
algebra for the operators $(\widehat{D}_a, \widehat{F}_{ab})$ is
simply a gauge transformation of the Snyder algebra for the
operators $(\widehat{D}^{(0)}_a, \widehat{Q}_{ab})$. Therefore the
resulting geometry determined by the vector fields
\eq{mass-vector} will not be changed and the constraint
\eq{de-sitter} will be preserved. So the coordinate change in
terms of flat connections should be a Killing symmetry of the
background geometry \eq{metric-snyder} as was explained in
Eq.\eq{inverse-emergent-metric} and correspond to a global Lorentz
transformation in higher dimensions, which was precisely the case
for two-dimensional geometries. For example, from Eq.\eq{de-sitter}
or Eq.\eq{ind-snyder}, one can deduce that $x^A
\to x^{\prime A} = {\Lambda^A}_B x^B$ where ${\Lambda^A}_B \in
SO(5-p,p)$.

We observed that a higher dimensional manifold in general emerges
from a NC $\star$-algebra $\CA_\theta$ defined by a Poisson
structure rather than a symplectic structure. Another notable
difference from the two-dimensional case is that the underlying
action \eq{mass-ikkt} contains fluctuations by non-flat connections
and so nontrivial metric deformations. This means that the action
\eq{mass-ikkt} describes a fluctuating geometry, not a rigid
geometry. The Snyder algebra \eq{snyder-gauge} clearly shows that
the action \eq{mass-ncu1} allows such fluctuations by non-flat
connections as an on-shell solution. Indeed the algebra
\eq{snyder-gauge} can be understood as the Lorentz algebra
\eq{high-lorentz} after the identification $M_{AB} = i \kappa^2
\widehat{F}_{AB} = i \kappa^2 (\widehat{F}_{ab}, \, \widehat{F}_{d+1,a}
\equiv - \frac{i}{\sqrt{\kappa}} \widehat{D}_a)$.

Suppose that the fluctuations \eq{d-exp} in commutative limit are described by smooth functions
$z^a(x) = x^a + \kappa  A^a(x)$ where $x^a$ describe the vacuum
geometry in Eq.\eq{metric-snyder}. Then one can map the solution
$D_a = g_{ab} z^b/\kappa \in C^\infty(M)$ to vector fields in
$\Gamma(TM)$ according to Eq.\eq{mass-vector}. Let us denote the
resulting vector fields as $(V_a, S_{ab} = [V_a,V_b])$ which satisfy
the Snyder algebra as easily inferred from Eq.\eq{snyder-gauge}. The
resulting Snyder algebra can be lifted to the Lorentz algebra in
five dimensions given by
\be \la{deform-lorentz}
S_{AB} = \kappa \Big( z_B \frac{\partial}{\partial z^A} - z_A
\frac{\partial}{\partial z^B} \Big)
\ee
where $V_a \equiv S_{5,a}/\sqrt{\kappa}$ and $z^A = z^A(x)$ are five-dimensional
coordinates satisfying $g_{AB} z^A z^B = (-1)^\sharp R^2$. Following
the same procedure as Eqs.\eq{metric-snyder} and \eq{ind-snyder},
the metric of fluctuating surface $M$ can be derived as\footnote{We
feel some remarks are necessary to correctly understand
Eq.\eq{f-metric-snyder} and to avoid any confusion. The equivalence
principle in general relativity guarantees that there always exists
a locally inertial frame at an arbitrary point $P$ in spacetime
where the metric becomes locally flat, i.e., $ds^2|_P =
\eta_{\alpha\beta} d \xi^\alpha d\xi^\beta$. But the local inertial frame
$\xi^\alpha = \xi^\alpha(x)$ is valid only on a local coordinate
patch and cannot be globally extended over all spacetime unless the
spacetime is flat. Similarly, Eq.\eq{snyder-gauge} implies that it
is always possible to choose a local coordinate $z^a$ such that the
metric at $P$ locally looks like the background geometry
\eq{metric-snyder}. But we have to notice that the local coordinates
$z^a (x) = x^a + \kappa A^a(x)$ depend on dynamical gauge fields
satisfying the equations of motion or the Snyder algebra
\eq{snyder-gauge} and so should not be regarded as a globally
constant curvature spacetime as if the local inertial frame does not
mean a flat spacetime.}
\bea \la{f-metric-snyder}
ds^2 &=& \mathfrak{G}_{ab} dz^a \otimes dz^b \xx &=& \Big(g_{ab} +
g_{55} \frac{z_a z_b}{z_5^2} \Big) dz^a \otimes dz^b = \Big(g_{ab} +
g_{55} \frac{z_a z_b}{z_5^2} \Big) \frac{\partial z^a}{\partial x^c}
\frac{\partial z^b}{\partial x^d} dx^c \otimes dx^d \xx
&=&  \Big(g_{ab} + g_{55} \frac{x_a x_b}{x_5^2} \Big) dx^a
\otimes dx^b + \big({\rm deformations \; of \;} \CO(A) \big)
\eea
and
\bea \la{flat-snyder}
ds^2 &=& \Big(g_{ab} + g_{55} \frac{z_a z_b}{z_5^2} \Big) dz^a
\otimes dz^b \xx
&=& g_{AB} \frac{\partial z^A}{\partial x^a} \frac{\partial
z^B}{\partial x^b} dx^a \otimes dx^b = g_{AB} dz^A
\otimes dz^B.
\eea
If the solution \eq{d-exp} is understood as a general coordinate
transformation $x^A \mapsto z^A = z^A(x)$ in $(d+1)$ dimensions, one
may notice that Eq.\eq{flat-snyder} is certainly a higher
dimensional analogue of the two-dimensional result \eq{co-metric}.

Now let us recapitulate why the emergent geometry we have examined so
far is completely consistent with all the rationale inferred from
the algebraic point of view. We are interested in the emergent
geometry derived from the mass-deformed IKKT matrix model
\eq{mass-ikkt}. We observed that the equations of motion can be
derived from the Snyder algebra \eq{snyder-alg}. An essential point
is that the Snyder algebra \eq{snyder-alg} in $d$ dimensions can be
lifted to the $(d+1)$-dimensional Lorentz algebra
\eq{high-lorentz}. So the $d$-dimensional Snyder algebra can be represented by
the $(d+1)$-dimensional Lorentz generators with the constraint
$g_{AB} z^A z^B = (-1)^\sharp R^2$. As we know, the Lorentz algebra
\eq{high-lorentz} represents a global symmetry of
$(d+1)$-dimensional flat spacetime. Therefore the emergent gravity
determined by the Snyder algebra
\eq{snyder-alg} can always be embedded into $(d+1)$-dimensional {\it
flat} spacetime although the $d$-dimensional geometry is highly
nontrivial. From the $d$-dimensional point of view, the geometry of
hypersurface $M$ is emergent from dynamical gauge fields as
the map \eq{mass-vector} definitely implies. One may clearly see this
picture from Eq.\eq{flat-snyder}. First recall that $z^a(x) = x^a +
\kappa A^a(x)$ where $A^a(x)$ describe fluctuations around the
background spacetime whose metric is given by Eq.\eq{metric-snyder}.
But the last result of Eq.\eq{flat-snyder} shows that the dynamical
fluctuations of the manifold $M$ can again be embedded into the
$(d+1)$-dimensional flat spacetime, but its embedding is now
described by the ``dynamical" coordinates $z^A (x) = x^A +
\kappa A^A(x)$.

Like the two-dimensional case, one may consider a nonlinear
deformation of the Snyder algebra by replacing the mass term in the
action
\eq{mass-ikkt} by a general polynomial as follows:
\be \la{gen-mass-ikkt}
S_G = \Tr \big( \frac{1}{4} M_{ab} M^{ab} -
\frac{1}{2} M_{ab} [X^a, X^b] + \frac{\kappa}{2}
G(X) \big).
\ee
Then the equations of motion \eq{eom2-matrix} are replaced by
\be \la{gen-eom2}
[M^{ab}, X_b] + \kappa \Big[ \frac{\partial G(X)}{\partial X_a} \Big] = 0
\ee
where the second term is a formal expression of the matrix ordering
under the trace $\Tr$ as Eq.\eq{eom-general}. Equation \eq{gen-eom2} could
be derived by considering the nonlinear version of the Snyder
algebra
\eq{snyder-alg}
\be \la{nl-snyder}
[X^a, M^{bc}]  = \kappa f^{abcd}  \Big[ \frac{\partial G(X)}{\partial X_d}
\Big]
\ee
where $f^{abcd} = g^{ac} g^{bd} - g^{ab} g^{cd}$ has been chosen to
recover the linear Snyder algebra with $G(X) = (d-1) \kappa X_a X^a$. As
long as the polynomial $G(X)$ is explicitly given, the commutator
$[M^{ab}, M^{cd}]$ can be calculated by applying the Jacobi identity
\be \la{nl-snyder3}
[M^{ab}, M^{cd}] = [ M^{ab}, [X^c, X^d]] =  [[M^{ab}, X^c], X^d] -
[[M^{ab}, X^d], X^c]
\ee
and using the algebra \eq{nl-snyder}. The right-hand side of
Eq.\eq{nl-snyder3} can eventually be arranged into the form
$\kappa G^{ac}(X) M^{bd} + \cdots $ using the commutation relation
\eq{nl-snyder}. Therefore, the nonlinear deformation of the Snyder
algebra described by the action \eq{gen-mass-ikkt} seems to work. So
it will be interesting to investigate whether the nonlinear Snyder
algebra can still have a higher dimensional interpretation like the
linear case and what kind of vacuum geometry arises from a given
polynomial $G(X)$.

\section{Discussion and Conclusion}

Here we discuss the fact that the constant curvature space described by the
Snyder algebra \eq{snyder-alg} can be represented as a coset space
$G/H$. In other words, the $d$-dimensional hypersurface $M$ is a
homogeneous space. To be specific, we have the following coset
realization of $M$:
\bea \la{coset}
&& {\bf S}^d = SO(d+1)/SO(d), \xx && dS_d = SO(d,1)/SO(d-1,1),
\\ && AdS_d = SO(d-1,2)/SO(d-1,1). \nonumber
\eea
Taking $G$ to be a Lie group as in Eq.\eq{coset}, the coset manifold
endows a Riemannian structure as we already know. Split the Lie
algebra of $G$ as $\IG = \IH \oplus \IK$ where $\IH$ is the Lie
algebra of $H$ and $\IK$ contains the coset generators. The
structure constants of $G$ are defined by \ct{coset}
\bea \la{coset-lie}
&& [H_i, H_j] = f_{ij}^k H_k, \qquad H_i \in \IH, \xx && [H_i, K_a]
= f_{ia}^j H_j + f_{ia}^b K_b,  \qquad K_a \in \IK, \\
&& [K_a, K_b] = f_{ab}^i H_i + f_{ab}^c K_c. \nonumber
\eea
If $f_{ia}^j = 0$, the coset space $G/H$ is said to be reductive
and, if $f_{ab}^c = 0$, it is called symmetric.

In order to realize the coset space \eq{coset} from the Snyder
algebra \eq{snyder-alg}, it is obvious how to identify the
generators in $\IH$ and $\IK$: $K_a = i X_a \in \IK$ and $H_i =
M_{ab} \in \IH$. From this identification, we see that the coset
space \eq{coset} is symmetric as well as reductive, which is a
well-known fact. Therefore it will be interesting to see how the
emergent geometry from the Snyder algebra \eq{snyder-alg} can be
constructed from the Riemannian geometry of the coset space $G/H$.
The whole geometry of $G/H$ can be constructed in terms of coset
representatives
\be \la{coset-rep}
L(y) = e^{y^a K_a}, \qquad (a=1, \cdots, {\rm dim}G - {\rm dim}H)
\ee
where the local coordinates $y^a$ parameterize the coset $gH$ for
any $g \in G$. Under left multiplication by a generic element $g$ of
$G$, the coset representative \eq{coset-rep} will be transformed to
an another representative $L(y')$ of the form
\be \la{coset-tr}
g L(y) = L(y') h, \qquad h \in H,
\ee
where $y'$ and $h$ depend on $y$ and $g$ and on the way of choosing
representatives.

Consider the Lie algebra valued one-form
\be \la{lie-1-from}
V(y) = L^{-1}(y) d L(y) = V^a(y) K_a + \Omega^i(y) H_i.
\ee
The one-form $V^a(y) = V^a_\mu (y) dy^\mu$ is a covariant frame
(vielbein) on $G/H$ and $\Omega^i_\mu (y) dy^\mu$ is called the
$H$-connection. Under left multiplication by a constant $g \in G$,
the one-form \eq{lie-1-from} transforms according to
Eq.\eq{coset-tr} as
\be \la{1-form-tr}
V(y') = h L^{-1}(y) g^{-1} d \big( g L(y) h^{-1} \big) = h V(y)
h^{-1} + h d h^{-1}.
\ee
One can check using Eq.\eq{1-form-tr} that the left action of $G$ on
$V^a(y)$ is equivalent to an $SO(d)$ or $SO(d-1,1)$ rotation on
$V^a(y) \;(d = {\rm dim} G/H)$ \ct{coset}. The metric of the coset
space $G/H$ can be written in terms of the vielbeins in
Eq.\eq{lie-1-from} as
\be \la{coset-metric}
\mathfrak{G}^{(0)}_{\mu\nu} (y) = g_{ab} V^a_\mu (y) V^b_\nu (y)
\ee
where $g_{ab}$ is the flat coset metric and the metric
\eq{coset-metric} is invariant under the left action of $G$ due to
the property \eq{1-form-tr}.

Bause the metric \eq{metric-snyder} describes the coset manifolds
\eq{coset}, it will be equivalent to the $G$-invariant metric
\eq{coset-metric}. Note that the metric \eq{metric-snyder} is also $G$-invariant
as Eq.\eq{ind-snyder} definitely shows. So let us check the
Riemannian structure of the coset spaces \eq{coset}. The
differential properties of the one-form \eq{lie-1-from} are
expressed by the Maurer-Cartan equation
\be \la{mc-eq}
dV + V \wedge V = 0.
\ee
Using Eq.\eq{coset-lie}, one can decompose the Maurer-Cartan
equation \eq{mc-eq} as
\bea \la{mc-decom-1}
&& dV^a + \frac{1}{2} f_{bc}^a V^b \wedge V^c + f_{ib}^a
\Omega^i \wedge V^b = 0, \\
\la{mc-decom-2}
&& d\Omega^i + \frac{1}{2} f_{ab}^i V^a \wedge V^b + f_{ja}^i
\Omega^j \wedge V^a + \frac{1}{2} f_{jk}^i \Omega^j \wedge
\Omega^k = 0.
\eea
In our case the above equations are much simpler because $f_{bc}^a =
f_{ja}^i = 0$. Combining Eq.\eq{mc-decom-1} together with the
torsion free condition $DV^a = dV^a + {\omega^a}_b \wedge V^b = 0$
yields the spin connection on $G/H$
\be \la{spin-conn}
{\omega^a}_b = f_{ib}^a \Omega^i.
\ee
The Riemann curvature tensor is defined in term of ${\omega^a}_b$ by
\be \la{riemann-curv}
{R^a}_b = d {\omega^a}_b + {\omega^a}_c \wedge {\omega^c}_b.
\ee
Substituting \eq{spin-conn} into \eq{riemann-curv} and using
Eq.\eq{mc-decom-2} lead to the curvature tensors
\bea \la{curvature-tensor}
{R^a}_b &=& - \half f_{ib}^a f^i_{cd} V^c \wedge V^d +
\big(f_{ic}^a f_{jb}^c - \half f_{ij}^k f_{kb}^a \big) \Omega^i
\wedge \Omega^j \\ &\equiv&
\half {R^a}_{bcd} V^c \wedge V^d \nonumber
\eea
where the second term in Eq.\eq{curvature-tensor} vanishes because
of the Jacobi identity $[[K_a,H_i],H_j] + [[H_i,H_j], K_a] + [[H_j,
K_a], H_i] = 0$.

Comparing the coset algebra \eq{coset-lie} with the Snyder algebra
\eq{snyder-alg} leads to the identification of the structure
constant $f_{id}^a = g^{ae} f_{eid}$ for $i=[bc]$
\be \la{structure-const}
f_{a[bc]d} = f_{abcd} = g_{ac} g_{bd} - g_{ab} g_{cd}.
\ee
Then the Riemann curvature tensor \eq{curvature-tensor} of coset
manifold $G/H$ is given by\footnote{According to the identification
\eq{high-lorentz}, the first Snyder algebra for four-dimensional anti-de Sitter space
is given by $[X^a, X^b] = \kappa [M^{5,a}, M^{5,b}] = \kappa g^{55}
M^{ab}$. Thus the anti-de Sitter space will be equally cared by the
replacement $f^i_{ab} \to g_{55}f^i_{ab}$ in the algebra
\eq{coset-lie}. That is the reason why the $g_{55}$ factor
comes in Eq.\eq{riemann-tensor}.}
\be \la{riemann-tensor}
R_{abcd} = - f_{aib} f^i_{cd} = - f_{aefb} {f^{ef}}_{cd} =
g_{55}(g_{ac} g_{bd} - g_{ad} g_{bc}).
\ee
As was shown in Eq.\eq{metric-snyder}, a vacuum geometry of the
Snyder algebra \eq{snyder-alg} is also given by an Einstein manifold
of constant curvature and is precisely the same as
Eq.\eq{riemann-tensor}. Therefore, we confirm that the vacuum
geometry of the Snyder algebra \eq{snyder-alg} is described by the
$G$-invariant metric \eq{coset-metric} of the coset space $G/H$. But
we have to notice that the Snyder algebra \eq{snyder-gauge} is
in general defined by dynamical gauge fields fluctuating around the vacuum
manifold $G/H$ as Eq.\eq{f-metric-snyder} clearly shows. One might
already notice that the generators in Eq.\eq{coset-lie} are constant
matrices while those in Eq.\eq{snyder-alg} are in general mapped
to NC fields in $\CA_\theta$ as in Eq.\eq{snyder-gauge}. Therefore,
it should be interesting to directly derive Einstein's equations \ct{hsyang}
to incorporate all possible deformations induced by gauge fields from the Snyder algebra,
whose metric may be $G$-invariant as always, as we checked in Eq.\eq{flat-snyder}.
We hope to address this issue in the near future.

Let us conclude with some remarks about the significance of emergent
geometry based on the results we have obtained. The emergence usually
means the arising of novel and coherent structures, patterns and
properties through the collective interactions of more fundamental
entities, for example, the superconductivity in condensed matter
system or the organization of life in biology. In our case, we are
talking about the emergence of a much more bizarre object: gravity. A
stringent point of emergent gravity is to require that spacetime
should also be emergent simultaneously according to the picture of
general relativity.

What does the emergence of spacetime mean ? It means that the
emergent gravity should necessarily be background independent where
the prior existence of any spacetime structure is not {\it a priori}
assumed but should be defined by fundamental ingredients in
quantum gravity theory. We have already exhibited such examples with
the matrix actions \eq{ikkt}, \eq{master-action} and \eq{mass-ikkt}.

Let us pick up the simplest example \eq{mass-cs} to illuminate how
some geometry emerges from a background independent theory. Note
that the action \eq{mass-cs} is a ``zero-dimensional" matrix model. In
order to define the action \eq{mass-cs}, we did not introduce any
kind of spacetime structure. We only have three Hermitian matrices
(as objects) which are subject to the algebraic relations
\eq{eom-su2} and \eq{casimir} (as morphisms).\footnote{Indeed
$g_{AB}$ is nothing more than a symbol for the algebraic
characterization of ``zero-dimensional" matrices although it will be realized
as a three-dimensional metric in the end.} From these algebraic relations between objects,
we can derive a geometry by mapping the matrix algebra to a Poisson algebra
or a NC $\star$-algebra, as was shown in Sec. 2.
Depending on the choice of an algebraic relation characterized by the signature of $g_{AB}$,
we get a different geometry. The underlying argument should be familiar, in particular,
with the representation theory of Lie groups and Lie algebras.

A profound aspect of emergent geometry is that a background-independent formulation
can be realized with matrix models, as we
illustrated with the actions \eq{ikkt}, \eq{master-action} and
\eq{mass-ikkt}. In this approach, an operator algebra, e.g.,
$\star$-algebra defined by NC gauge fields, defines a relational
fabric between NC gauge fields, whose prototype at a macroscopic world
emerges as a smooth spacetime geometry. In this scheme, the geometry is a derived concept
defined by the algebra. One has to specify an underlying algebra to
talk about a corresponding geometry. Furthermore, a smooth
geometry is doomed in a deep NC space, whereas an algebra between objects plays a more
fundamental role. Therefore, the motto of emergent
gravity is that an algebra defines a geometry.

As we observed in Eq.\eq{poisson-lie}, the map between a Poisson
algebra $(C^\infty(M), \{-,-\}_\theta)$ and the Lie algebra
$(\Gamma(TM), [-,-])$ of vector fields is a Lie algebra
homomorphism. This means that a geometric structure determined by
the Lie algebra $(\Gamma(TM), [-,-])$ is faithfully inherited from
the Poisson algebra $(C^\infty(M), \{-,-\}_\theta)$. Thus the
map between an underlying algebra and its emergent geometry should
be structure-preserving, i.e., a homomorphism. This homomorphism
is also true even for a general Poisson structure. Actually it
should be required for consistency of emergent gravity. If not, one
could not say that a geometry can be derived from an algebra.

In our case, this implies that an algebraic structure in a matrix
theory will be encoded in a geometric structure of emergent gravity.
Note, as we showed in Sec. 3, the maximally symmetric spaces
in Eq.\eq{coset} can be derived from the Snyder algebra
\eq{snyder-alg} by applying the map \eq{mass-vector}. And recall that
those $d$-dimensional symmetric spaces can always be embedded in a
$(d+1)$-dimensional flat spacetime. If so, a natural question is how
this geometric property is encoded in the Snyder algebra
\eq{snyder-alg}. As Eq.\eq{high-lorentz} shows, the geometric
property is precisely realized as the fact that the $d$-dimensional
Snyder algebra can be arranged into the Lorentz algebra in $(d +
1)$-dimensional flat spacetime. Although the equivalence between
the $d$-dimensional Snyder algebra and the $(d + 1)$-dimensional
Lorentz algebra is a well-known fact, it is a nice nontrivial
check that the algebraic structure of the Snyder algebra has been
consistently encoded in the geometric property of emergent spacetime
since the emergent gravity has to respect the homomorphism from an
algebra to a geometry for consistency.

As a completely different direction, we may consider the matrices
$(X^a, M^{ab})$ as independent dynamical coordinates, which satisfy
the $(d+1)$-dimensional Lorentz algebra \eq{high-lorentz}.
As an example, a three-dimensional sphere ${\bf S}^3$ appears in this way from the
$SU(2)$ algebra \eq{eom-su2} as we discussed in the footnote
\ref{s2s3}. In this case there are $d(d+1)/2$ coordinates in total and
so we will get some $d(d+1)/2$-dimensional manifold from the algebra
\eq{snyder-alg} or \eq{high-lorentz}. Although we do not know what the underlying
Poisson structure is in this case, we guess that the resulting emergent
geometry derived from the Lorentz algebra \eq{high-lorentz} would be
a group manifold of $SO(d+1-p, p)$ as can be inferred from the
three-dimensional case. To clarify this issue will be an interesting
future work.

\section*{Acknowledgments}

We thank V. Rivelles for initial collaboration and discussions. HSY
thanks Kuerak Chung and Kimyeong Lee for helpful discussions. MS
thanks Bum-Hoon Lee for the invitation to the Center for Quantum Spacetime,
Seoul; FAPESP for the visit to the University of S\~ ao Paulo, where part of
the work was done and DST (India) for support in the form of a
project. The work of H.S. Yang was supported by the RP-Grant 2009 of
Ewha Womans University.

\appendix

\section{Two-dimensional Snyder Algebra}

Here we will show that the two-dimensional version of the Snyder algebra \eq{snyder-alg} is
precisely equal to the three-dimensional $SO(3-p,p)$ Lie algebra \eq{eom-su2}.

In two dimensions, the Snyder algebra \eq{snyder-alg} reads as
\be \la{2-snyder}
[X^1, X^2] = M^{12}, \quad [X^1, M^{12}] = - \kappa g^{11} X^2,
\quad [X^2, M^{12}] = \kappa g^{22} X^1.
\ee
If one defines $M^{12} \equiv \pm i \lambda X^3
\;(= -i \lambda {\varepsilon^{12}}_3 X^3)$,
one can immediately see that the Snyder algebra \eq{2-snyder} can be written as
the form of the Lie algebra \eq{eom-su2} with $\kappa = - \lambda^2 \det g_{AB}$.
Conversely, if one defines $X^A \equiv \frac{i}{2 \lambda} {\varepsilon^A}_{BC} M^{BC}
\;(A,B,C = 1,2,3)$, the Snyder algebra \eq{2-snyder} takes the form of the three-dimensional
Lorentz algebra \eq{high-lorentz}. Note that the two-dimensional Snyder algebra \eq{2-snyder}
is the equation of motion derived from the action \eq{mass-ikkt}, which can be rewritten as
the action \eq{mass-cs} for the three-dimensional Lie algebra with the above identification.
It might be remarked that the three dimensions is special in the sense that
an antisymmetric rank-2 tensor is dual to a vector, i.e., $M^{AB} = - i \lambda {\varepsilon^{AB}}_C
X^C$ and so the Lorentz algebra \eq{high-lorentz} can be expressed as the form \eq{eom-su2}
only in three dimensions.

As we discussed in Sec. 2, the quadratic form $C_2 \equiv \sum_{A,B=1}^3 g_{AB} X^A X^B$ is a
Casimir invariant of $SO(3-p, p)$ Lie algebra, i.e.,
\be \la{casimir-comm}
[X^A, C_2] = 0, \qquad \forall A = 1,2,3.
\ee
Because $X^A = \frac{i}{2 \lambda} {\varepsilon^A}_{BC} M^{BC}$, Eq.\eq{casimir-comm} can be
rewritten as
\be \la{lorentz-casimir-comm}
[M^{AB}, C_2] = 0, \qquad \forall M^{AB} \in SO(3-p, p).
\ee
This means that $C_2$ is a Lorentz invariant, which can also be derived using the commutation
relation
\be \la{3-lorentz}
[X^A, M^{BC}] = \kappa \Big(g^{AC} X^B - g^{AB} X^C \Big).
\ee
The invariance \eq{casimir-comm} implies that $C_2$ is a multiple of the identity element
of the algebra such as Eq.\eq{casimir}. From the viewpoint \eq{lorentz-casimir-comm},
$C_2$ is an invariant under $SO(3-p,p)$ Lorentz transformations.
Therefore the Casimir invariant \eq{casimir} can simultaneously be interpreted as
a Lorentz invariant which reduces to the three-dimensional version of Eq.\eq{de-sitter}, i.e. $\sum_{A,B=1}^3 g_{AB} x^A x^B = (-1)^\sharp R^2$, in a classical limit.

In summary, it was shown that the three-dimensional $SO(3-p,p)$ Lie algebra \eq{eom-su2}
is isomorphic to the two-dimensional version of the Snyder algebra \eq{snyder-alg} where
the embedding condition \eq{de-sitter} can be identified with the Casimir invariant \eq{casimir}.



\end{document}